\documentclass[fleqn,usenatbib]{mnras}

\usepackage{newtxtext,newtxmath}

\usepackage[T1]{fontenc}

\DeclareRobustCommand{\VAN}[3]{#2}
\let\VANthebibliography\thebibliography
\def\thebibliography{\DeclareRobustCommand{\VAN}[3]{##3}\VANthebibliography}

\usepackage{graphicx}	
\usepackage{amsmath}	
\usepackage{xcolor}

\definecolor{coinblue}{HTML}{244EA3}

\hypersetup{
    colorlinks=true,
    linkcolor=coinblue,
    citecolor=coinblue,
    urlcolor=coinblue
}

\title[SAGUI]{\includegraphics[scale=0.0175]{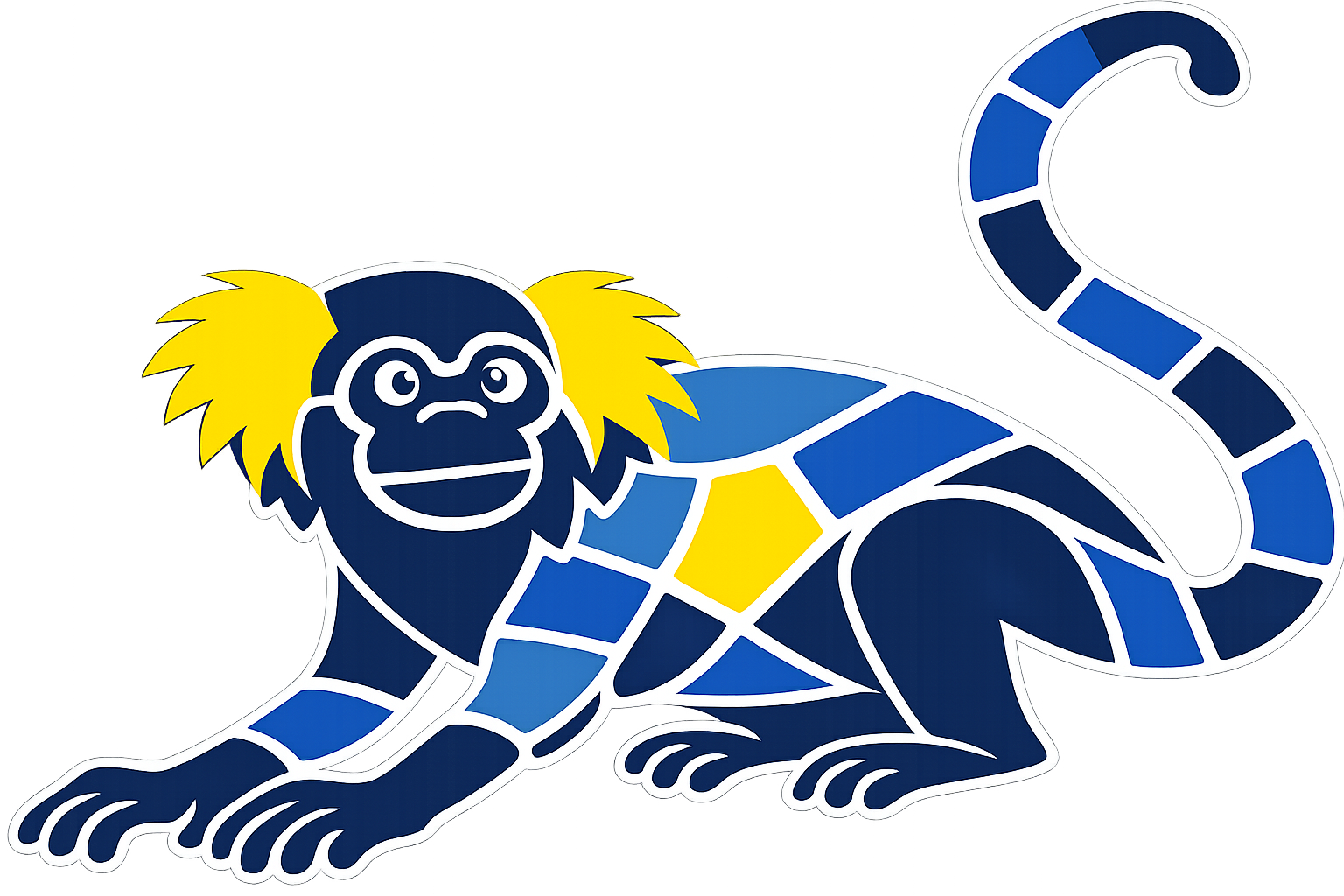}
SAGUI: SED-based Segmentation of Multi-band Galaxy Images — Application to JADES in GOODS-South}

\author[de Souza et al.]{Rafael S. de Souza,$^{1,2,3}$\thanks{E-mail: rd23aag@herts.ac.uk}
\thanks{drsouza@ad.unc.edu}
Andressa Wille,$^{2}$
Shravya Shenoy,$^{1}$
Aarya A. Patil,$^{4}$
Alberto Krone-Martins,$^{5}$
\newauthor
Ana L. Chies-Santos,$^{2}$
Celine Boehm,$^{6}$
Reinaldo R. Rosa,$^{7}$
Thallis Pessi,$^{8}$
Emille E. O. Ishida,$^{9}$
\newauthor
Kristen C. Dage,$^{10}$
Lilianne Nakazono,$^{11}$
Phelipe Darc,$^{12}$
Rupesh Durgesh,$^{13}$
for the COIN collaboration
\\
$^{1}$Centre for Astrophysics Research, University of Hertfordshire, Hatfield AL10 9AB, UK\\
$^{2}$Instituto de Física, Universidade Federal do Rio Grande do Sul, Porto Alegre, RS 90040-060, Brazil\\
$^{3}$Department of Physics \& Astronomy, University of North Carolina at Chapel Hill, NC 27599-3255, USA\\
$^{4}$Max-Planck-Institut für Astronomie, Königstuhl 17, D-69117 Heidelberg, Germany\\
$^{5}$Donald Bren School of Information and Computer Sciences, University of California, Irvine, CA 92697, USA\\
$^{6}$School of Physics, The University of Sydney and ARC Centre of Excellence for Dark Matter Particle Physics, NSW 2006, Australia\\
$^{7}$Lab for Computing and Applied Mathematics, COPDT-INPE-MCTI, São José dos Campos, SP 12245-010, Brazil\\
$^{8}$European Southern Observatory, Alonso de Córdova 3107, Vitacura, Casilla 19001, Santiago, Chile\\
$^{9}$Université Clermont Auvergne, CNRS/IN2P3, LPC, F-63000 Clermont-Ferrand, France\\
$^{10}$ International Centre for Radio Astronomy Research—Curtin University, GPO Box U1987, Perth, WA 6845, Australia\\
$^{11}$Observatório Nacional / MCTI, Rua General José Cristino 77, Rio de Janeiro, RJ, 20921-400, Brazil\\
$^{12}$ Centro Brasileiro de Pesquisas F\'isicas, Rua Xavier Sigaud, 150, Urca, Rio de Janeiro, Brazil\\
$^{13}$Independent researcher, Germany
}

\date{Accepted XXX. Received YYY; in original form ZZZ}

\pubyear{\the\year{}}

\begin{document}
\label{firstpage}
\pagerange{\pageref{firstpage}--\pageref{lastpage}}
\maketitle

\begin{abstract}
We present {\sc sagui}, a modular framework for the analysis of multi-band imaging data in spatially resolved galaxies, with synergies to integral-field spectroscopy (IFS). Building on the spectro-spatial paradigm introduced by {\sc capivara} for IFS data, {\sc sagui} extends this approach to imaging datasets, enabling a coherent, pixel-level treatment of spatial and spectral information across multiple bands. The method follows a two-stage strategy: a starlet-based decomposition is first used to identify and mask spatial structures across multiple scales while suppressing noise, and a spectral-similarity analysis then partitions the image into coherent pixel groups that preserve spectral consistency. In addition to compact and high-contrast structures, the framework incorporates a dedicated statistical treatment, based on a copula transform, to identify and recover faint, diffuse low-surface-brightness components.
We demonstrate the method across a diverse range of galaxy morphologies, highlighting its ability to characterize complex spatial structures, including clumps, bars, interacting systems, and low-surface-brightness features. As a case study, we apply it to eleven morphologically diverse galaxies from the \textit{James Webb Space Telescope} Advanced Deep Extragalactic Survey (JADES) in the GOODS--South field. {\sc sagui} is released under an MIT license and is available at \href{https://rafaelsdesouza.github.io/sagui/}{GitHub}.
\end{abstract}




\section{Introduction}
The 2020s have witnessed a milestone in imaging and photometric studies in optical and near-infrared astronomy. The wealth of exquisite, high-quality data observed by the James Webb Space Telescope has been nothing short of remarkable. Moreover, this year marks the start of the long-awaited Legacy Survey of Space and Time \citep[LSST;][]{Ivezic2019} at the Vera C. Rubin Observatory. Together with forthcoming and ongoing facilities such as \textit{Euclid} \citep{EuclidCollaboration2022}, the Chinese Space Station Survey Telescope  \citep{2026SCPMA}, and the \textit{Nancy Grace Roman Space Telescope} \citep{Akeson2019}, these surveys will establish a transformative foundation for statistical studies of galaxy structure and  substructures, including bars, bulges, halos, discs and pseudo-components \citep{Binney2008,Simard2011}. They will also prove invaluable for low-surface brightness science \citep{Englert25}. These facilities deliver multi-filter imaging across broad optical and near-infrared wavelength ranges. Such data can be viewed as low-dimensional hyperspectral cubes, where each pixel encodes a spectral energy distribution (SED).

Prior to the widespread availability of multi-band imaging surveys and integral field spectroscopy, studies of galaxy evolution relied on integrated-light measurements or central single-fiber spectra from surveys such as the Sloan Digital Sky Survey \citep{York2000} and Galaxy and Mass Assembly \citep[GAMA;][]{Driver2009}. Although these data established canonical scaling relations, they lacked spatial resolution, and the inferred stellar masses and star formation rates (SFRs) depend sensitively on aperture size. In particular, at lower redshifts, the effect of a fixed aperture becomes more significant; for instance, in late-type spiral galaxies, where substantial star formation occurs in the outer discs, this can  lead to underestimated SFRs. At the same time, contamination from nuclear emission can result in overestimated SFRs (e.g. \citealt{2016richards_SAMI}). Integrated measurements may underestimate stellar masses due to the dominance of young luminous populations (“outshining” effects; \citealt{Sorba2015}), reinforcing the need for spatially resolved analyses. 

Integral-field spectroscopy (IFS) transformed the field by enabling resolved maps of stellar populations, ionised gas, and kinematics. Surveys including SAURON \citep{deZeeuw02}, ATLAS\textsuperscript{3D} \citep{Cappellari2011}, CALIFA \citep{Sanchez2012}, MaNGA \citep{Bundy2015}, Hector \citep{Bryant2016},  FORNAX3D \citep{sarzi18}, SINS/zC-SINF \citep{Forster2018}, KMOS3D \citep{Wisnioski2019}, SAMI \citep{SAMIDR3_2021}, and  WEAVE \citep{2024Jin_WEAVE} have provided spatially resolved insights into galaxy assembly and dynamical evolution. In parallel, the Multi Unit Spectroscopic Explorer \citep[MUSE;][]{Bacon2010}, has enabled a broad range of targeted and survey-style galaxy programmes, including PHANGS-MUSE \citep{Emsellem2022}, MAGPI \citep{Foster2021}, GECKOS \citep{geckos25}, and MAUVE \citep{attwater26}, among others.

In parallel, spatially resolved SED modelling \citep{Abdurro2018} has shown that multi-band photometry can recover stellar ages, dust attenuation, and structural gradients when spectroscopy is unavailable. Simulation-based studies further demonstrate that galaxy properties obtained by summing pixel-level measurements are broadly consistent with those from integrated photometry, particularly at spatial resolutions of $\gtrsim$ 1 kpc \citep[e.g.][]{2018Smith}. Observational results support this picture, with resolved SED analyses reproducing coherent scaling relations such as the spatially resolved star formation main sequence \citep{Abdurro2017} and mitigating biases inherent to unresolved measurements \citep{Sorba2018}.

Despite these advances, important methodological limitations remain, particularly in observational studies. Pixel-level SED fitting is highly sensitive to noise, while binning strategies such as Voronoi tessellation \citep{Voronoi1908,Franz91} can improve the signal-to-noise (S/N) ratio \citep{Cappellari2003} at the potential expense of spectral coherence, thereby fragmenting physically meaningful structures.
The challenge is therefore not merely statistical efficiency but astrophysical awareness: segmentation must ideally be a trade-off between  physically meaningful components and morphological structure. Computer vision offers a relevant analogy. Semantic segmentation in hyperspectral imaging has proven effective in remote sensing and robotics \citep{Medellin2023}. Astronomical datasets share similar characteristics — high dimensionality, heterogeneous noise, and strong spatial correlations — but introduce additional constraints, including extreme dynamic range, dust attenuation, and the coexistence of emission-line and continuum features. These properties require domain-aware segmentation strategies.

The need for such approaches becomes evident when considering the physical complexity of galaxies. Stellar systems comprise multiple structural components—discs, spiral arms, bulges, bars, clumps, and nuclear clusters—embedded within a multiphase interstellar medium. Each exhibits distinct age, metallicity, and kinematic signatures shaped by internal processes and external perturbations throughout  cosmic time. Spiral arms and bars not only define the galactic morphology but also regulate angular momentum transport and secular evolution \citep[e.g.][]{sellwood2014}. Interactions ranging from minor fly-bys to major mergers redistribute baryons, producing tidal tails, bridges, and shells while triggering enhanced star formation \citep[e.g.][]{luo2014, ferreira2025}.
Compact starburst regions \citep[e.g.][]{calabro2019, elmegreen2021, he2022} and disturbed morphologies therefore act as tracers of dynamical evolution. Sub-galactic analyses are essential for connecting spatially localized processes to global transformation \citep{Helmi2020}. Early HST observations revealed clumpy kiloparsec-scale structures \citep[e.g.][]{cowie1995, elmegreen2005, elmegreen2009}, and the increased sensitivity of \textit{JWST} now enables detailed studies of star formation on even smaller physical scales \citep[e.g.][]{claeyssens2023, mowla24, messa24, claeyssens26, benotto26}. Spatially and spectrally coherent segmentation is therefore not a cosmetic refinement, but a prerequisite for linking localized physical processes to galaxy-wide evolution.

Building on our IFS-based spectral segmentation framework \texttt{capivara} \citep{capivara2025}, we introduce \texttt{sagui}, which extends this approach to multi-band photometric imaging. While preserving the core principle of clustering spatial pixels in SED space, \texttt{sagui} incorporates modifications required for the photometric regime, including multiscale starlet-based denoising \citep{starck_murtagh_2006, Melchior2018} and computational refinements to distance-matrix operations suited to higher-resolution imaging data,  including improved handling of missing data. Spatially resolved SED-fitting frameworks such as \texttt{piXedfit} \citep{piXedfit2021} and related photometric tools have demonstrated the scientific power of spatially resolved stellar population modelling. In practice, these analyses frequently rely on Voronoi binning or similar schemes to stabilize fits in low S/N regimes  \citep[e.g.][]{Fetherolf2020,Tan2022,Thaina2026}. Although such strategies improve statistical robustness, they may compromise spectral homogeneity or fragment physically coherent structures.

Our aim is not to replace these frameworks, but to complement them by addressing the upstream segmentation problem: identifying spatially coherent and spectrally homogeneous regions, i.e., regions with similar SED shapes, prior to population synthesis modelling (SED-fitting). In this sense, \texttt{sagui} complements source-detection and deblending tools by focusing on SED-coherent region segmentation within resolved galaxy light. The method is guided by two principles: (1) defining a physically motivated spatial envelope; and (2) clustering pixels in SED space.
Together, these components yield regions that enforce spectral consistency, enabling robust inference of stellar populations, dust content, and star-formation activity without relying on S/N–driven binning schemes that may blur spectrally distinct populations. 

This paper is structured as follows: Section~\ref{sec:data} describes the dataset and its motivation; Section~\ref{sec:method} presents the methodology; Section~\ref{sec:results} discusses the main results, including spatially resolved maps of the physical properties of our sample after segmentation, as well as a case study based on a copula transform \citep{nelsen_2006} combined with the segmentation to investigate low-surface-brightness structures. Section~\ref{sec:conclusions} summarises our conclusions. Throughout this paper, we assume a flat $\Lambda$CDM cosmology with $H_0 = 70$ km s$^{-1}$ Mpc$^{-1}$, $\Omega_m = 0.3$, and $\Omega_\Lambda = 0.7$.

\section{Data and sample selection}
\label{sec:data}

Throughout this paper, we analyse broadband imaging data to assess the performance of the \texttt{sagui} segmentation framework. Owing to the broad wavelength coverage of NIRCam imaging in the GOODS-South field we select ten galaxies in the main sample, plus one low-surface-brightness case study, observed with the \textit{James Webb Space Telescope} as part of the Advanced Deep Extragalactic Survey \citep[JADES;][]{rieke2023,eisenstein2023a,eisenstein2023b}. The sample is constructed to span a range of clear morphological substructures—such as star-forming clumps, bars, spiral arms, and merger signatures—providing a representative test bed for segmentation.
To ensure adequate spatial resolution while maintaining morphological diversity, the selected systems are restricted to $z \lesssim 1$ (see Table~\ref{tab:galaxies} and Section~\ref{sec:data}), although a formal quantification of the method’s redshift limit has yet to be performed.

\begin{figure*}
  \centering
  \includegraphics[width=\linewidth]{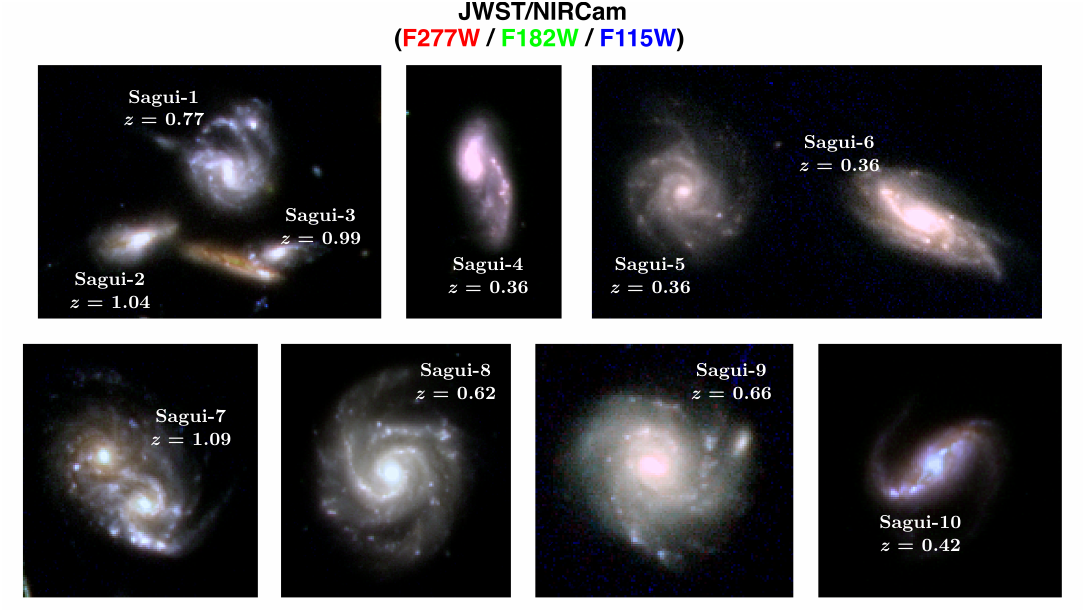}
  \caption{\textit{JWST}/NIRCam color composites of the galaxies analyzed in this study. Their identifiers and redshifts are indicated in each panel. The images were constructed using the filters F277W (red), F182M (green) and F115W (blue).}
  \label{fig:jades}
\end{figure*}

Figure~\ref{fig:jades} shows NIRCam colour composites constructed from F277W (R), F182M (G), and F115W (B) filters. Background levels are estimated and subtracted using low-percentile sky values, and intensities are scaled using a common non-linear (asinh) stretch derived from the luminance distribution of each system. This approach preserves faint tidal features and clumpy structure while preventing saturation in bright cores. A mild colour normalisation was applied to balance channel intensities. 

Sagui-1, Sagui-2, and Sagui-3 lie close in projection but differ in redshift, providing a test of algorithmic behaviour in crowded fields, with Sagui-3 potentially representing a merging system. Sagui-4 exhibits a strongly disturbed, post-merger morphology, with compact clumps distributed along an extended tidal feature, possibly associated with recent star formation. Sagui-5 and Sagui-6 lie at the same redshift but differ structurally: the former displays a disc-like morphology with visible spiral structure, while the latter appears elongated, with stellar emission stretched along a preferred axis, suggestive of ongoing interaction. Sagui-7 presents a disturbed system with two bright central components embedded in a diffuse envelope, consistent with a near-coalescence phase. Sagui-8 and Sagui-9 are relatively undisturbed, face-on spirals rich in star-forming clumps, while Sagui-10 hosts a prominent bar together with visible star-forming knots.

We use \textit{JWST}/NIRCam imaging in fourteen broad and medium bands (F090W, F115W, F150W, F182M, F200W, F210M, F277W, F335M, F356W, F410M, F430M, F444W, F460M, and F480M), spanning roughly $0.8-5 \mu$m in wavelength coverage. However, due to variations in the spatial coverage of individual filters, particularly the medium bands, not all galaxies of the sample are observed in fourteen filters. As a result, the number of available filters for each galaxy varies from 8 to 14 filters, with most galaxies covered in the full set of bands.

For each galaxy, the images were aligned and cropped to a common field of view across all filters before stacking into a multi-band cube with dimensions ($x$, $y$, filter). The spatial extent ($x$, $y$) of each cube was defined to encompass the full extent of the target, depending on the morphology and environment of each system (e.g. isolated galaxies or interacting pairs). The rectangular cutouts have typical sizes ranging from $\sim$150 to $\sim$300 pixels per side, corresponding to $\sim$0.1–0.2 arcmin.
All images share a pixel scale of 0.04 arcsec pix$^{-1}$, corresponding to the scale of the long-wavelength NIRCam mosaics.
Because the NIRCam point spread function (PSF) varies significantly with wavelength, direct comparison of pixel-level fluxes across bands would introduce artificial colour gradients and bias SED-based clustering. We therefore homogenised the imaging to a common effective angular resolution prior to segmentation, by matching the PSF of all bands to that of the F444W filter (FWHM $\simeq 0.145$ arcsec). Empirical PSFs for the JADES GOODS--South field, measured by \citet{delavega2025}, were used to construct convolution kernels for each band. All data processing was performed using the \textsc{lightstack} code \footnote{https://github.com/AndressaWille/lightstack}, which implements the steps described above.

We obtain spectroscopic redshifts primarily from the MUSE Hubble Ultra Deep Field Survey \citep{2017A&A...608A...1B, 2023A&A...670A...4B}, except for Sagui-5, for which we use a VIMOS spectrum \citep{2009A&A...494..443P}. We determine redshifts by fitting Gaussian profiles to prominent nebular emission lines (e.g.\ [O,\textsc{ii}] $\lambda3727$, H$\beta$, [O,\textsc{iii}] $\lambda5007$, H$\alpha$). We fit each line independently, and centroid offsets relative to rest-frame wavelengths yield individual redshift estimates. When multiple lines are available, we compute a weighted mean using inverse-variance weights based on the centroid uncertainties. The final redshift values, reported to four significant digits, reflect the typical precision ($\sigma_z \sim 10^{-4}$) provided by the MUSE spectral resolution. Table~\ref{tab:galaxies} lists coordinates and spectroscopic redshifts of the galaxies.

\begin{table}
\centering
\caption{Coordinates and spectroscopic redshifts of the eleven galaxies in our sample. Ten of these systems are shown in Figure~\ref{fig:jades} and form the main analysis sample, while Sagui-11 is discussed separately as a low-surface-brightness case study in Section \ref{sec:lsb}.}
\begin{tabular}{lccc}
\hline
Galaxy ID & RA (J2000) & DEC (J2000) & Redshift \\
\hline
Sagui-1 & 03:32:36.50 & $-27$:46:29.5 & 0.7649 \\
Sagui-2 & 03:32:36.40 & $-27$:46:31.6 & 1.0385 \\
Sagui-3 & 03:32:36.60 & $-27$:46:31.2 & 0.9965 \\
Sagui-4 & 03:32:34.20 & $-27$:45:54.4 & 0.3664 \\
Sagui-5 & 03:32:36.38 & $-27$:51:18.0 & 0.3583 \\
Sagui-6 & 03:32:35.95 & $-27$:51:18.6 & 0.3583 \\
Sagui-7 & 03:32:35.59 & $-27$:46:26.8 & 1.0878 \\
Sagui-8 & 03:32:40.87 & $-27$:46:16.8 & 0.6222 \\
Sagui-9 & 03:32:13.70 & $-27$:49:34.5 & 0.6651 \\
Sagui-10 & 03:32:39.27 & $-27$:45:33.0 & 0.4148 \\
Sagui-11 & 03:32:19.50 & $-27$:52:17.89 & 1.0930 \\
\hline
\end{tabular}
\label{tab:galaxies}
\end{table}

\section{Methodology}
\label{sec:method}
The segmentation strategy implemented in \texttt{sagui} builds upon the spectral-clustering philosophy originally introduced in \texttt{capivara} \citep{capivara2025}, but adapts it to multi-band imaging data. While \texttt{capivara} was designed for IFS cubes with high-resolution spectra per spaxel, \texttt{sagui} operates on broadband and medium-band photometric images, where each pixel carries an SED. In addition, \texttt{sagui} incorporates an explicit morphology-preserving spatial masking stage based on undecimated wavelet analysis, which is not part of the original \texttt{capivara} implementation. The method therefore consists of the following, conceptually distinct steps: i) morphology-aware spatial masking, based on a starlet decomposition of a white-light image; ii) spectral segmentation, via hierarchical clustering of pixel SEDs. Additionally, we experiment with a copula-based transform to improve sensitivity to diffuse, low–surface-brightness (LSB) outskirts; this extension is described in Section~\ref{sec:lsb}.

\begin{figure*}
\centering
 \includegraphics[width=\linewidth]{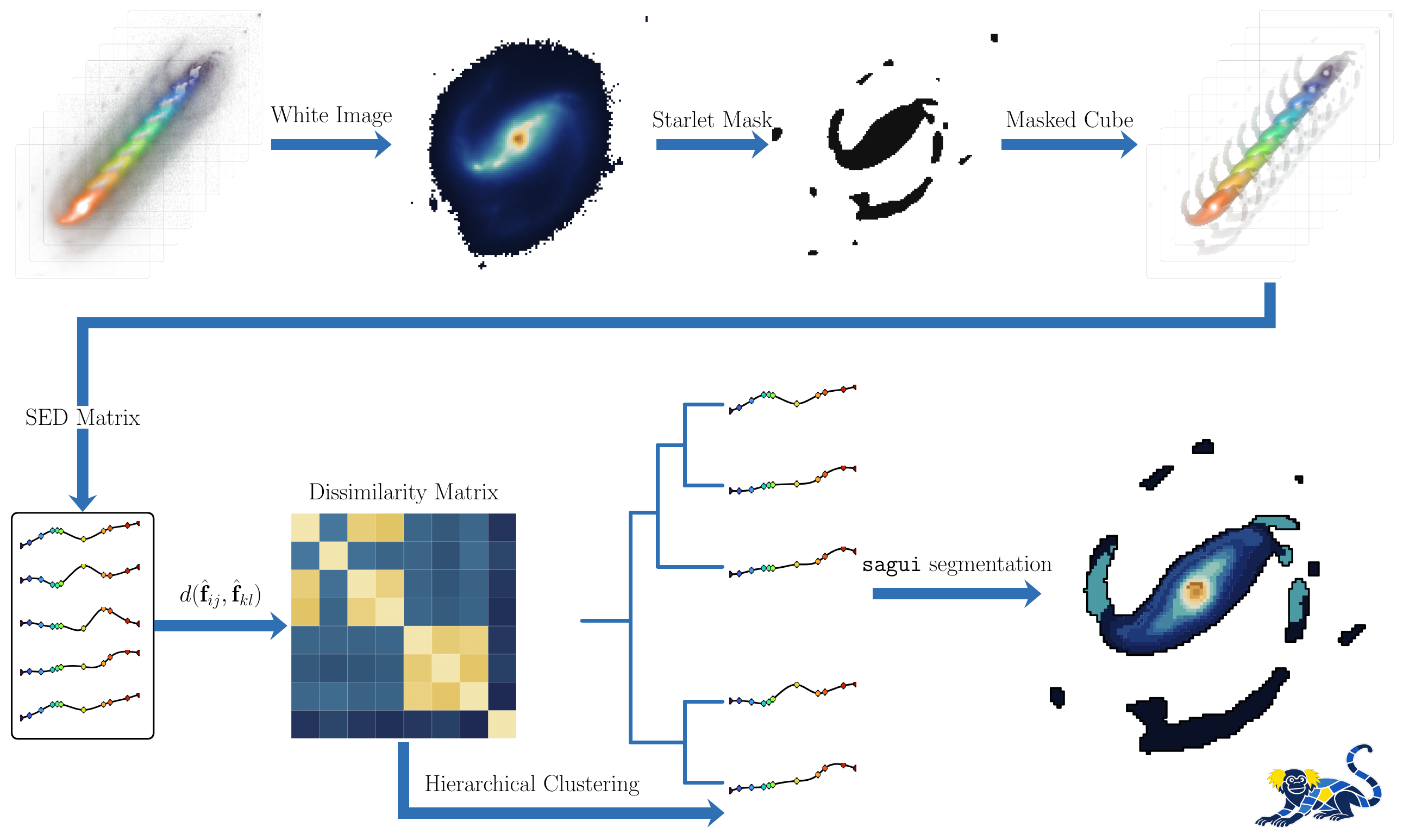}  
\caption{\texttt{sagui} workflow illustration. Starting from a multi-band imaging cube, a white-light image is collapsed and used to define a starlet-based foreground support. Pixel SEDs within this support are then compared in photometric space, and hierarchical clustering is applied to the corresponding dissimilarity matrix. The final cluster assignments are projected back onto the image plane, yielding a spatial segmentation whose regions follow coherent galactic structures while preserving similarity in SED shape.}
\label{fig:workflow}
\end{figure*}

Figure~\ref{fig:workflow} shows a schematic overview of how \texttt{sagui} operates. Starting from the multi-band image cube, we first construct a white-light image and use its starlet decomposition to define a foreground support that preserves the main galaxy morphology while suppressing background-dominated pixels. The pixel SEDs within this support are then compared through a pairwise spectral dissimilarity matrix, and hierarchical clustering is applied to identify groups of pixels with similar photometric behaviour. Finally, the resulting labels are mapped back onto the image plane, yielding a SED-based spatial segmentation.

\subsection{Starlet Mask}

Defining the spatial extent of a galaxy is intrinsically challenging, since galaxies do not have sharp physical edges and their surface-brightness profiles fade gradually into the background. A variety of methods have been proposed for this task, ranging from noise-based detection techniques such as \texttt{noisechisel} \citep{Akhlaghi2015} to U-Net-based deep-learning approaches \citep{Iglesias2024}. Here we employ an isotropic undecimated wavelet transform, commonly referred to as the \emph{starlet transform} \citep[e.g.][]{starck_murtagh_2006,starck_fadili_murtagh_2007}. Because the starlet transform is shift-invariant and defined directly on the original pixel grid, structures identified at different spatial scales remain co-registered, avoiding artefacts associated with downsampling or pixel shifts. This is particularly advantageous for astronomical imaging, where galaxies exhibit features over a wide range of spatial scales and where no explicit geometric prior, such as symmetry or connectivity, need be imposed. The resulting multiscale representation is used to define a morphology-preserving foreground mask that isolates the main galaxy structure while excluding most background-dominated pixels. We implemented a customised version of this transform within \texttt{sagui}; the corresponding functionality will also be incorporated into a future release of \texttt{capivara}.

\begin{figure*}
\centering
\includegraphics[width=\linewidth]{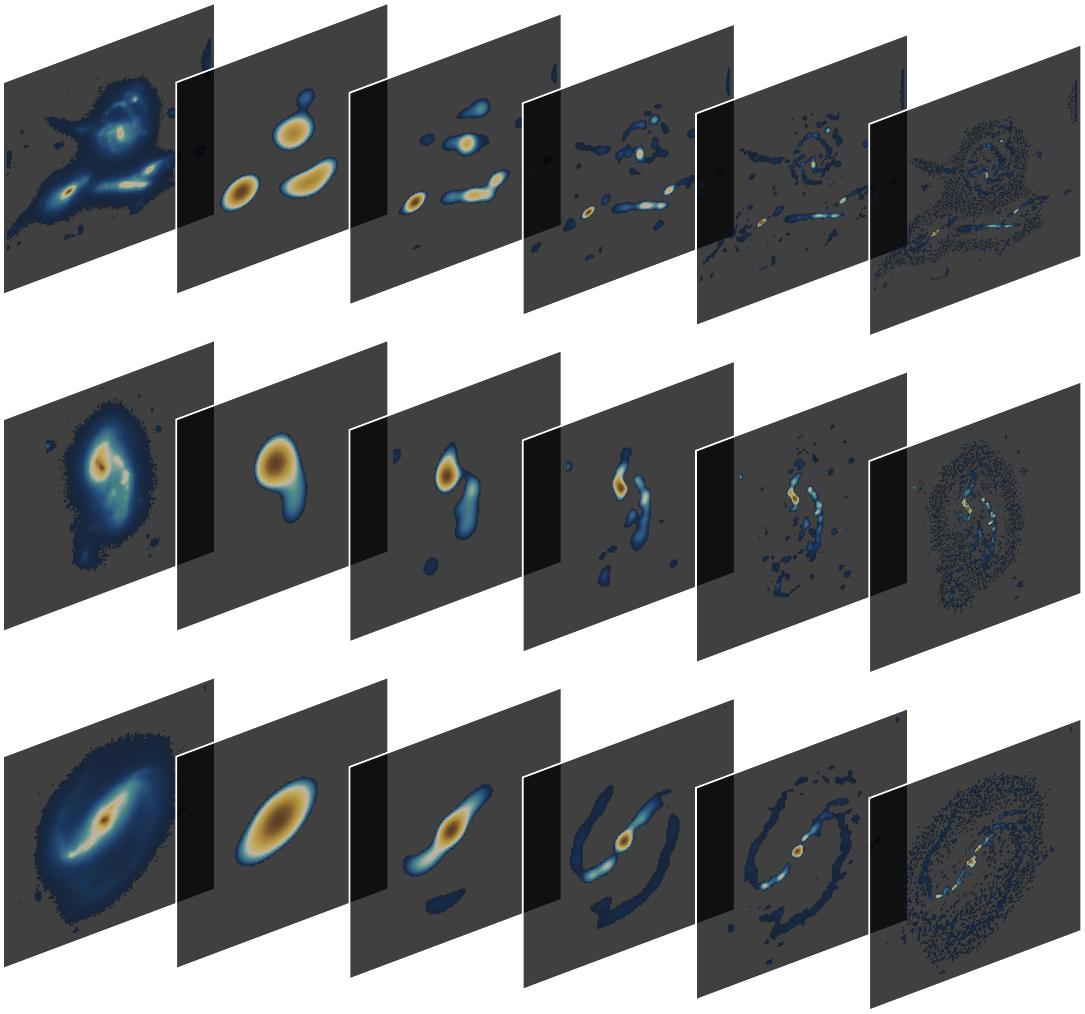}
\caption{Starlet decomposition for representative galaxies (Sagui-1–3, Sagui-4, and Sagui-10). In each case, the original image is exactly recovered by summing the detail coefficients across all five scales and the coarse component. The finest scale ($j=1$) is dominated by pixel-scale fluctuations, while intermediate scales trace coherent galactic structure. Larger scales isolate progressively smoother and more extended components.}
\label{fig:starlet_sagui123_10}
\end{figure*}

For the masking step, we first collapse the multi-band cube over all available bands to construct a white-light image,
\begin{equation}
I_0(x,y) = \sum_{b=1}^{N_{\mathrm{band}}} I_b(x,y),
\end{equation}
where \(I_b(x,y)\) denotes the image intensity in band \(b\).
The starlet transform then iteratively constructs a sequence of progressively smoothed images \(\{c_j\}_{j=1}^{J}\) by convolution with a dilated B$_3$-spline scaling filter,
\begin{equation}
h[k] = \tfrac{1}{16}(1,4,6,4,1),
\end{equation}
where dilation at scale $j$ is achieved by inserting $2^{j-1}-1$ zeros
between filter coefficients.
The parameter \(J\) sets how far the multiscale decomposition is allowed to
probe toward large spatial scales. In the \`a trous construction this happens
dyadically: each successive scale samples the same B$_3$-spline filter at twice the previous pixel spacing. Thus, scale \(j\) is sensitive to structures over
a characteristic width of order \(2^{j-1}\) pixels, with the full filter support
extending over \(4 \times 2^{j-1}+1\) pixels. Increasing \(J\) therefore allows
broader, smoother galaxy features to appear in the starlet coefficient maps
instead of being absorbed into the final coarse component \(c_J\). Conversely,
if \(J\) is too small, extended low-surface-brightness emission can remain in
\(c_J\) and may not contribute to the foreground support used for the mask.
At each scale, with \(c_0 \equiv I_0\), the wavelet coefficients are defined by differencing successive smooth components,
\begin{equation}
w_j = c_{j-1} - c_j .
\end{equation}
The decomposition therefore yields a hierarchy of detail maps $\{w_j\}$ capturing progressively larger spatial structures, together with a final coarse component $c_J$. Owing to the undecimated construction, all components retain the original spatial dimensions, and the transform satisfies the exact reconstruction identity
\begin{equation}
I_0 = \sum_{j=1}^J w_j + c_J .
\end{equation}
Applied to the white-light image, this produces multiscale detail coefficients \(\{w_j(x,y)\}_{j=1}^{J}\) and a smooth residual component \(c_J(x,y)\), all defined on the native spatial grid. Figure~\ref{fig:starlet_sagui123_10} illustrates the resulting multiscale decomposition in practice.

To identify the main galaxy structure, we reconstruct a support image from the
intermediate starlet scales while excluding both the finest scale, \(j=1\), and
the coarse residual \(c_J\). The finest scale is typically dominated by
pixel-scale fluctuations, whereas the coarse residual mainly captures the
smoothest large-scale component of the image, including background variations.
For the fiducial segmentations we use \(J=5\) and keep scales
\(j=2,\ldots,5\), so that
\begin{equation}
R(x,y) = \sum_{j=2}^{5} w_j(x,y).
\end{equation}
This choice retains structures on scales larger than the pixel-scale noise
plane and up to the largest dyadic scale resolved by the adopted
decomposition, while leaving only the smoothest residual component in \(c_J\).
In practice, \(J\) should be chosen with respect to the image size, the PSF
width, and the angular extent of the diffuse structures of interest; here we
fix \(J=5\) for all galaxies to ensure a homogeneous foreground-mask
definition across the sample.
The foreground support is then obtained by binarising this reconstructed image,
\begin{equation}
M(x,y) =
\begin{cases}
1, & R(x,y)>0,\\
0, & R(x,y)\leq 0,
\end{cases}
\end{equation}
with non-finite pixels excluded. Since starlet coefficients encode local
contrast, positive values in the reconstructed support trace signal above the
local background.

\subsection{SED-based segmentation}

In contrast to geometric tessellations such as Voronoi binning \citep{Voronoi1908,Franz91}, which partition the image plane according to proximity, our approach partitions the feature space of pixel SEDs and maps the resulting clusters back onto the image grid. The resulting tessellation is therefore defined in SED space rather than on the image plane, in the spirit of feature-space partitioning in statistical pattern recognition \citep{bishop2006pattern}.
Each pixel $x_{ij}$ is represented by an SED over the available \textit{JWST}/NIRCam bands for that galaxy, with components $f_{ijb}$ given by the background-subtracted flux in band $b$.

Pixel similarity is quantified through distances between normalized SEDs. Let $\widehat{\mathbf f}_{ij} = (\widehat f_{ij1},\ldots,\widehat f_{ijB})$ denote the normalized SED vector of pixel $x_{ij}$. Before computing pairwise distances, each pixel SED is centred by subtracting
its own median flux across the available bands.
This normalization removes the zeropoint of each pixel SED and makes the clustering more sensitive to differences in SED shape than to absolute surface-brightness differences.
The distance between two pixels $x_{ij}$ and $x_{kl}$ is defined using the $\ell_p$ norm,
\begin{equation}
d(x_{ij},x_{kl}) =
\left\lVert
\widehat{\mathbf f}_{ij} - \widehat{\mathbf f}_{kl}
\right\rVert_p
=
\left(
\sum_{b=1}^{B}
\left|
\widehat f_{ijb} - \widehat f_{klb}
\right|^p
\right)^{1/p},
\label{eq:lp_phot}
\end{equation}
where $b=1,\ldots,B$ indexes the photometric bands.
Throughout this work we adopt $p=2$, corresponding to the Euclidean distance. The resulting pairwise dissimilarities are then used as input to a hierarchical agglomerative clustering using Ward’s criterion \citep{Ward1963,MurtaghLegendre2014}. Ward’s linkage is adopted because it merges clusters in a way that minimises the increase in within-cluster variance, making it particularly appropriate for identifying groups of pixels with similar SEDs. In contrast to more local linkage criteria, it generally yields regions that are more compact and more readily interpretable. The algorithm starts with each pixel treated as an individual cluster and then successively merges the pair of clusters that produces the smallest increase in total within-cluster variance. This iterative procedure yields a dendrogram that captures the hierarchical structure of SED similarity across the image. The final segmentation is obtained by cutting the dendrogram at the chosen number of clusters.

\subsection{Spatially Resolved SED fitting}

We estimate the physical properties of each segmented region using the \textsc{Prospector} SED fitting code \citep[][]{2021Johnson_Prospector}. The code employs the Flexible Stellar Population Synthesis (FSPS) model \citep{2009Conroy}, implemented through  \textsc{Python-fsps} \citep{2014PyFSPS}.  This method generates SED models that allow for a flexible treatment of both stellar populations and dust physics, while explicitly accounting for parameter uncertainties.  
Continuum and line emission for stellar populations are calculated using the nebular emission models implemented within FSPS (see \citealt{2017Byler_fspsmodels} for details) using the photoionization code \texttt{cloudy} \citep{1998Ferland_Cloudy}. Stellar mass is assigned a uniform prior over the range $10^{4.5}$–$10^{13.5}\,M_{\odot}$, spanning the stellar mass scales expected in our analysis, from individual star-forming regions to larger structures.
Stellar metallicity is allowed to vary uniformly between $10^{-2}\,Z_{\odot}$ and $10^{0.3}\,Z_{\odot}$. We adopt an initial mass function of \citet{2003Chabrier_IMF} and the \citet{2000Calzetti_DustAttenuation} attenuation law. 

For each region, we construct an integrated SED by summing the flux from all pixels contained within the segmentation mask in each photometric band. The total flux in each band is computed as the sum of the pixel fluxes, and uncertainties are propagated by combining the individual pixel errors in quadrature. 
The resulting multi-band photometry therefore represents the integrated light of the stellar population and dust content associated with each segment.
These segment-level SEDs are then fitted with \textsc{Prospector} using identical model assumptions and priors for all regions, thereby ensuring a uniform analysis. We adopt deliberately broad priors to encompass the expected spatial variations in galaxy properties; however, a more detailed, spatially resolved treatment will be explored in future work to better capture effects such as AGN contamination and other localized processes. Spectroscopic redshift estimates from Table \ref{tab:galaxies} are used, ensuring that the derived physical parameters are not biased by photometric redshift uncertainties.
For each galaxy, all segments are fitted at the same redshift. Internal kinematic shifts between regions are expected to be small compared with the width of the broadband and medium-band filters used here, but component-dependent redshifts could in principle be incorporated in future applications. 
Fixing the redshift also removes one degree of freedom from the model, which improves the accuracy of derived physical parameters and tightens their posterior distributions. We adopt an exponentially declining parametric star-formation history,
\begin{equation}
\mathrm{SFR}(t) \propto e^{-t/\tau},
\end{equation}
where $\tau$ is the characteristic time scale of the model. As shown by \citet{2019Carnall}, inferred galaxy properties, particularly SFRs, can vary by at least 0.3 dex depending on the assumed SFH.
In this work, however, we are primarily interested in relative variations among segmented regions rather than in absolute parameter values, so any systematic offsets introduced by the choice of SFH are expected to affect all regions in a broadly similar way and therefore not dominate the inferred spatial trends. Nevertheless, investigating the impact of alternative SFH priors would be valuable in future work, especially because non-parametric SFHs have been shown to improve the recovery of galaxy properties by reducing the outshining of older stellar populations by recent star formation \citep{2025Euclid} and by better capturing bursty star-formation episodes \citep{2024Haskell,Harvey2025}.
The posterior distributions returned by \textsc{Prospector} provide estimates of stellar mass, age, metallicity, star-formation rate, and dust attenuation. We compute the average star-formation rate over recent timescales of $\Delta t = 100$ Myr as
\begin{equation}
\mathrm{SFR}_{\Delta t}
=
\frac{1}{\Delta t}
\int_{t_{\mathrm{age}}-\Delta t}^{t_{\mathrm{age}}}
\mathrm{SFR}(t)\,dt,
\end{equation}
where $t_{\mathrm{age}}$ is the age of the stellar population. We also use the mass-weighted age,
\begin{equation}
\langle t \rangle_{M}
=
\frac{\int_{0}^{t_{\mathrm{age}}} t\,\mathrm{SFR}(t)\,dt}
{\int_{0}^{t_{\mathrm{age}}} \mathrm{SFR}(t)\,dt}.
\end{equation}
Uncertainties and parameter degeneracies are retained through the posterior sampling, allowing for robust error estimation on a segment-by-segment basis.\\
\\
The derived physical properties are mapped back to the spatial domain by assigning each segment’s median posterior values to all pixels within its segmentation mask. This yields two-dimensional maps of stellar mass surface density, star formation rate surface density, dust attenuation, and other inferred quantities, while preserving the underlying morphological segmentation (see Figures \ref{fig:starlet_sagui4}--\ref{fig:starlet_sagui10}).
In this way, segment-level SED fitting results are translated into spatially resolved maps that reflect the galaxy’s structural properties. These maps should be interpreted as region-based summaries projected onto the image plane, rather than independent pixel-level estimates. Their strength lies in combining the statistical robustness of region-integrated SED fitting with the preservation of spatial morphology.

\section{Application to selected JADES galaxies}
\label{sec:results}

\subsection{\texttt{Sagui} Segmentation}

\begin{figure*}
\centering
\includegraphics[width=\linewidth]{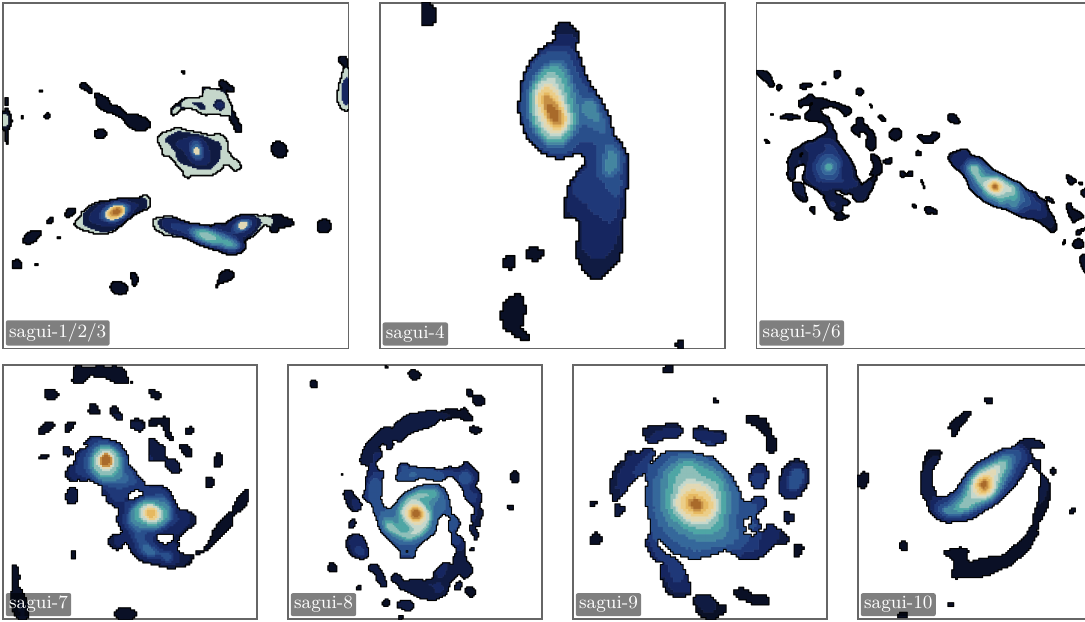}
\caption{Spatial distribution of the 20 segments detected by \texttt{sagui}. Different colours denote distinct segments.}
\label{fig:mosaic_segmentations}
\end{figure*}

We illustrate the workflow using a representative sample of eleven galaxies spanning a range of morphologies. Applying \texttt{sagui} to the multi-band images produces segmentation matrices that assign each pixel to a distinct SED-defined group. The resulting SED-based segmentations for 10 cases are shown in Figure~\ref{fig:mosaic_segmentations}. For this demonstration, we divide each galaxy into 20 segments, which offers a practical compromise between structural characterization and S/N ($\gtrsim 20$). We stress that the main purpose of this paper is methodological. A detailed astrophysical interpretation of the segmented regions is left to future work; the parameter choices adopted here are therefore intended to showcase the method’s capabilities rather than to deliver a comprehensive analysis of the galaxies themselves.

For each segmented region identified by \texttt{sagui}, we infer physical parameters using \textsc{Prospector} SED fitting and project the resulting estimates back onto the spatial grid. This yields two-dimensional maps of stellar population properties,  enabling a visual inspection of spatial variations in stellar mass, star-formation activity, metallicity, and dust attenuation. Because each segment corresponds to a coherent region in SED space, these maps provide a physically motivated partition of the galaxy, linking morphological structures to their underlying stellar populations. 

We emphasize that \texttt{sagui} is not designed to replace generic source-detection or deblending pipelines. Methods such as \textsc{SExtractor} \citep{Bertin1996}, \textsc{NoiseChisel} \citep{Akhlaghi2015}, and \textsc{scarlet} \citep{Melchior2018} are optimized for detecting, separating, and measuring sources in imaging data. The role of \texttt{sagui} is different: it operates after the construction of a spatially meaningful foreground mask and aims to partition galaxy light into regions that remain coherent in SED space, thereby providing a physically motivated input for resolved stellar-population analysis. In this sense, \texttt{sagui} is best understood as complementary to source-detection methods rather than as a competing alternative. The exact appearance of the segmentation depends on analysis choices such as the PSF homogenisation across bands, the starlet-mask parameters used to define the spatial envelope, and the adopted number of components. These parameters control the effective granularity of the partition and should therefore be adjusted to the scientific objective of the application.

\begin{figure}
\centering
\includegraphics[width=\linewidth]{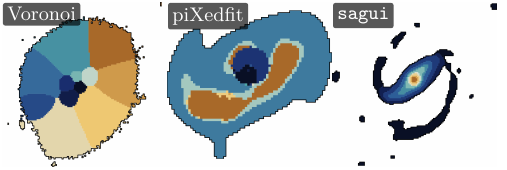}
\caption{Comparison of segmentation behaviour for Sagui-10 using 10 components. Left: pure Voronoi binning without a foreground mask. Middle: a \texttt{piXedfit}-like adaptive pixel-binning baseline, restricted to a \texttt{piXedfit}-like foreground mask. Right: \texttt{sagui}, which combines a starlet-derived foreground mask with clustering in SED space.}
\label{fig:seg_compare_sagui10}
\end{figure}

Figure~\ref{fig:seg_compare_sagui10} presents a direct comparison between three segmentation strategies applied to Sagui-10: pure Voronoi binning, Voronoi binning restricted to a \texttt{piXedfit}-like foreground mask, and \texttt{sagui}. This example is intended to separate the effects of the foreground support definition from those of the segmentation itself. Pure Voronoi binning provides a purely geometric partition driven by the target S/N, while the \texttt{piXedfit}-like variant introduces a more realistic foreground mask but retains the same geometric binning logic. By contrast, \texttt{sagui} combines a starlet-based support definition with SED-aware clustering. Visually, \texttt{sagui} better follows the bar, spiral arms, and central spheroidal component, whereas the geometric baselines tend to merge structures that are morphologically adjacent but photometrically distinct. This comparison highlights that the gain obtained with \texttt{sagui} is not only due to masking, but also to the way the segmentation uses multi-band information to preserve physically meaningful structures.

\begin{figure}
\centering
\includegraphics[width=\linewidth]{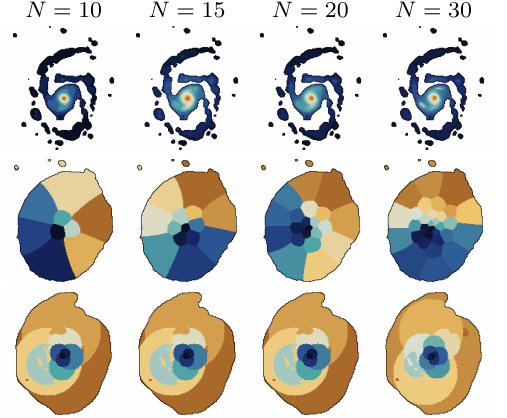}
\caption{Benchmark comparison for Sagui-8. Rows show \texttt{sagui}, pure
Voronoi binning, and the \texttt{piXedfit}-like baseline. Columns show nominal target group numbers \(N=10,15,20,30\).}
\label{fig:benchmark_sagui8}
\end{figure}

In Figure~\ref{fig:benchmark_sagui8} we compare \texttt{sagui} with two geometric binning baselines at different granularity levels for the spiral galaxy Sagui-8. We consider pure Voronoi tessellation and a Voronoi scheme restricted to a \texttt{piXedfit}-like foreground mask, and run both methods for nominal target numbers of regions \(N=\{10,15,20,30\}\). Because these
adaptive binning schemes are controlled by a S/N criterion rather than by fixing the exact number of bins, the recovered number of regions may differ slightly from the nominal target. We then run \texttt{sagui} with the
corresponding target number of components to enable a direct visual comparison between the segmentation strategies. In this regular spiral system, \texttt{sagui} more naturally follows the central concentration and the spiral-arm structure seen in the composite image, whereas the geometric tessellations tend to produce partitions that are less well aligned with the galaxy morphology. This distinction is important because our goal is not only to increase S/N, but also to partition the galaxy light into regions that remain coherent in SED
space.

\begin{figure}
\centering
\includegraphics[width=\linewidth]{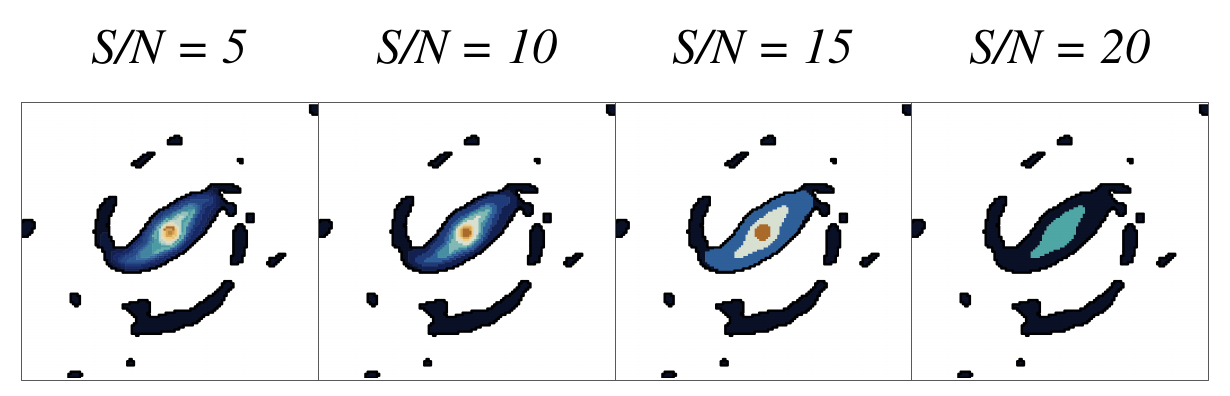}
\caption{Target-S/N segmentation test for Sagui-10. Each panel shows the \texttt{sagui} segmentation obtained when requiring a minimum regional signal-to-noise ratio of \(\mathrm{S/N}=5,10,15,\) and \(20\), respectively. Increasing the target S/N forces the algorithm to merge regions, producing progressively coarser segmentations while preserving the main photometric structures.}
\label{fig:sagui10_target_snr}
\end{figure}

Figure~\ref{fig:sagui10_target_snr} illustrates the complementary use of \texttt{sagui} when the segmentation granularity is set by a minimum regional S/N requirement rather than by a fixed number of components. For Sagui-10, increasing the target threshold from \(\mathrm{S/N}=5\) to $20$ progressively merges neighbouring regions, producing coarser maps with higher-quality integrated photometry. This provides a practical way to tune the segmentation for downstream SED fitting: lower thresholds preserve more spatial detail, while higher thresholds prioritise more precise flux measurements per region.

\begin{figure*}
\centering
\includegraphics[width=\linewidth]{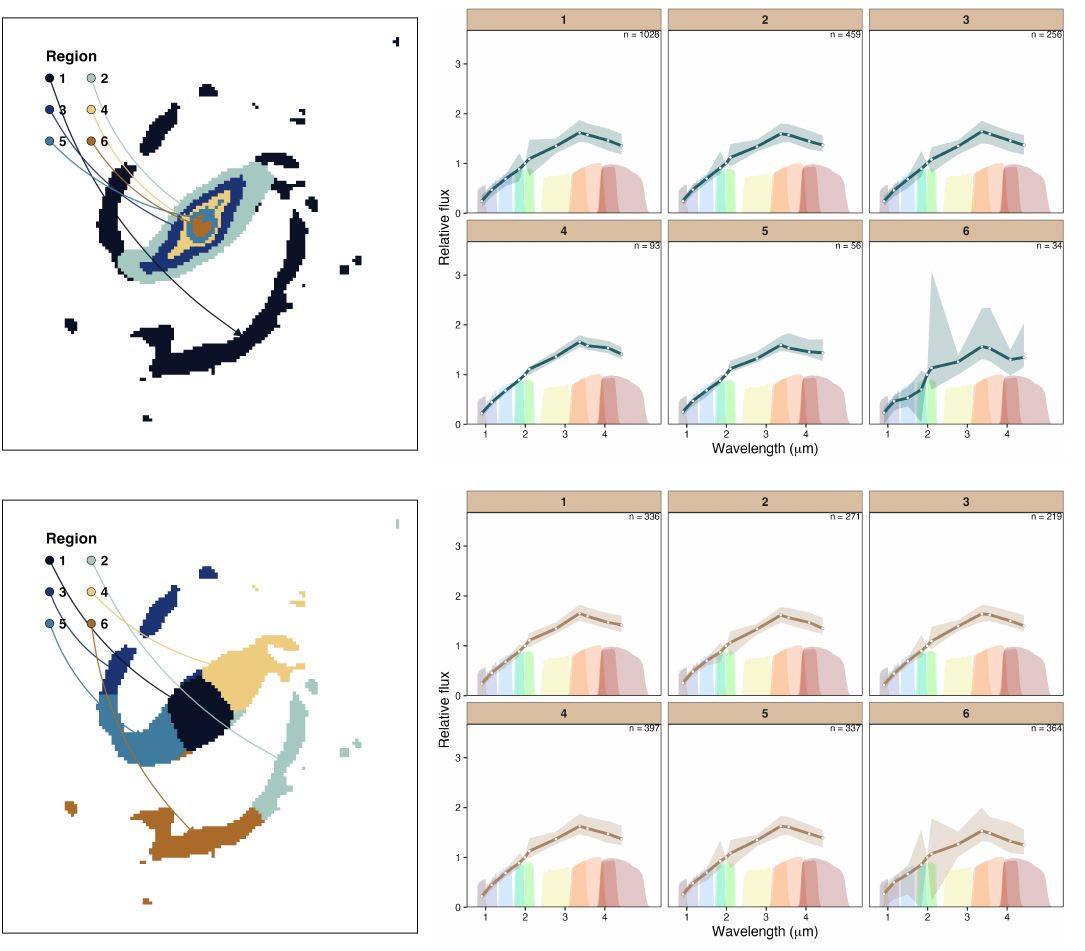}
\caption{Illustrative comparison for Sagui-10 using a shared starlet support, with \textsc{sagui} segmentation in the top panel and a Voronoi tessellation in the bottom panel, both using six regions. The left panels show the segmentation maps, while the right panels display bundles of normalized \textit{JWST} SEDs for the pixels in each segment. The shaded coloured bands indicate the approximate transmission windows of the \textit{JWST} filters.
}
\label{fig:sagui10_sed_bundle_compare}
\end{figure*}

Figure~\ref{fig:sagui10_sed_bundle_compare} shows a controlled comparison between \texttt{sagui} and Voronoi binning for Sagui-10, using the same starlet-derived mask and the same number of regions. The left panels show the segmentation maps, and the right panels the corresponding bundles of normalized \textit{JWST} SEDs for the pixels in each region. Because the mask is fixed, the differences reflect only the clustering step. The \texttt{sagui} segmentation follows the galaxy morphology in a non-parametric way, reproducing the bar and surrounding disc more naturally than the Voronoi partition. Since the clustering is hierarchical, the largest and most populated components are typically established first, with later refinements splitting the brighter inner structure. The resulting SED bundles are generally tighter than in the Voronoi case, indicating greater internal spectral homogeneity.

\begin{figure*}
\centering
\includegraphics[width=\linewidth]{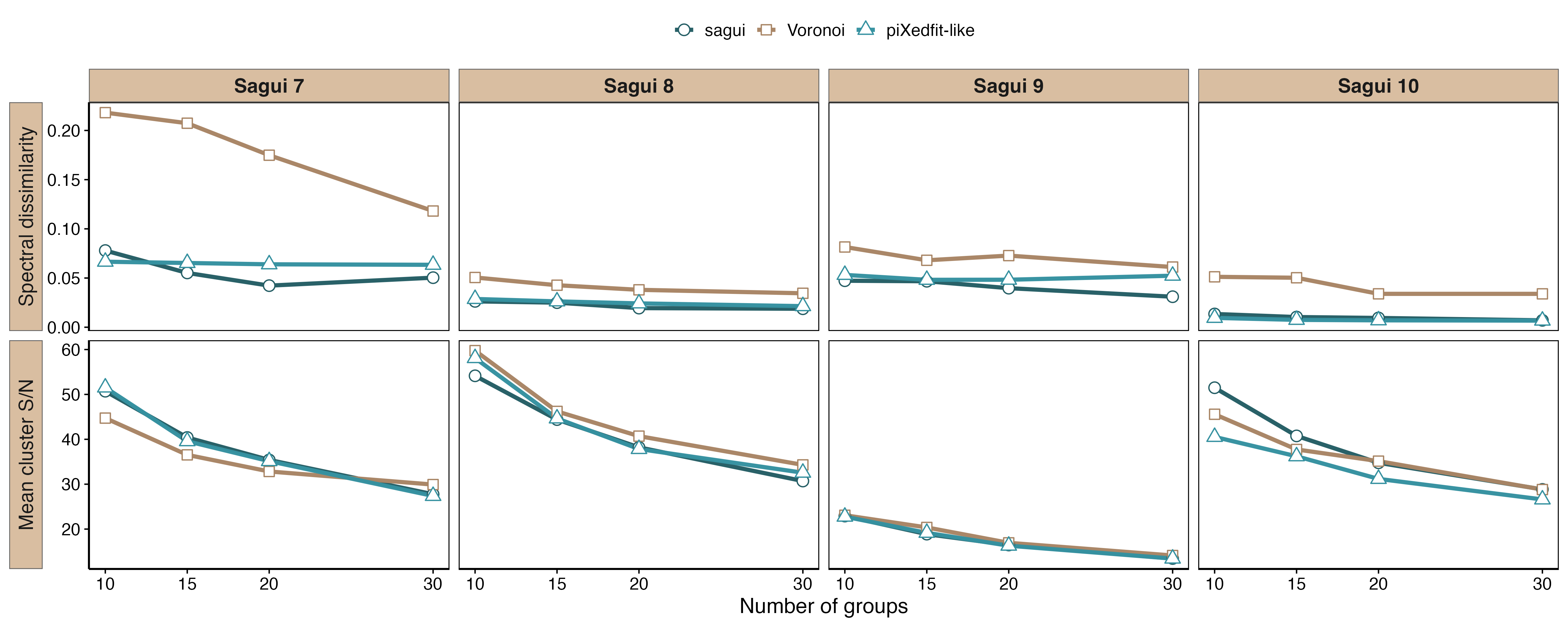}
\caption{Validation of the segmentation quality as a function of the number of groups for Sagui-7, Sagui-8, Sagui-9, and Sagui-10. The top row shows the spectral dissimilarity, while the bottom row shows the mean cluster S/N. Across all four galaxies, \texttt{sagui} yields lower spectral dissimilarity than Voronoi binning, indicating more spectrally coherent regions, while maintaining comparable or higher mean cluster S/N over most of the explored range.}
\label{fig:validation_by_ncomp}
\end{figure*}

While Figure~\ref{fig:sagui10_sed_bundle_compare} conveys a visual intuition about the behaviour of each tessellation method, it lacks a quantitative metric. In Figure~\ref{fig:validation_by_ncomp}, we introduce simple heuristics to assess segmentation performance as a function of the number of groups for four representative systems: Sagui-7, Sagui-8, Sagui-9, and Sagui-10. We compare \texttt{sagui}, Voronoi binning, and a \texttt{piXedfit}-like baseline using the spectral dissimilarity 
\((1 - \bar{\rho}_{\mathrm{cluster}})\),
where
\begin{equation}
\bar{\rho}_{\mathrm{cluster}} =
\frac{1}{K}\sum_{k=1}^{K}\bar{\rho}_{k},
\qquad
\bar{\rho}_{k} =
\frac{2}{n_k(n_k - 1)}
\sum_{\substack{i < j \\ i, j \in k}}
\mathrm{corr}\!\left(\tilde{f}_i, \tilde{f}_j\right),
\end{equation}
and the mean cluster S/N ratio
\begin{equation}
\overline{\mathrm{S/N}} =
\frac{1}{K}\sum_{k=1}^{K}
\left(\frac{\mathrm{S}}{\mathrm{N}}\right)_k.
\end{equation}
Here, \(K\) is the number of regions, \(n_k\) is the number of pixels in region \(k\), and  \(\tilde{f}_i\) is the scaled SED of pixel \(i\). In all galaxies, \texttt{sagui} yields systematically lower spectral dissimilarity than Voronoi binning, while maintaining comparable or higher mean cluster S/N.

\subsubsection{Sagui-1, -2, and -3}

This system comprises three galaxies, Sagui-1, Sagui-2, and Sagui-3, located in close projection within the JADES field but spanning distinct redshifts ($z = 0.77$, $z = 1.04$, and $z = 0.99$, respectively). Despite their apparent proximity on the sky, the galaxies are not physically associated, providing a representative example of a crowded high-redshift field in which multiple systems with overlapping isophotes and complex morphologies coexist. All three exhibit irregular and asymmetric light distributions, combining compact bright regions with extended low-surface-brightness features, making them well suited for evaluating the ability of the segmentation to disentangle projected structures while preserving morphological diversity.

Figure~\ref{fig:mosaic_segmentations} shows the spatial distribution of the 20 components identified by the \texttt{sagui} segmentation. Visual inspection reveals several distinct features. In Sagui-1 ($z=0.77$), the algorithm isolates a compact, high-SNR central component, surrounded by multiple asymmetric and elongated regions that trace extended stellar structures and diffuse emission. These features are suggestive of a disturbed morphology, potentially linked to interaction or  ongoing assembly, and are recovered without imposing any prior assumptions about galaxy symmetry or connectivity.
Sagui-2 ($z=1.04$) is segmented into a small number of relatively compact components, dominated by a bright core. The segmentation naturally separates these regions based on local surface-brightness coherence, capturing both the central concentration and fainter surrounding structures that would otherwise be blended in object-based approaches.
For Sagui-3 ($z=0.99$), the method identifies an elongated main component
and several secondary regions aligned with its major axis, consistent with a disc-like or tidally stretched morphology. Despite the projected
proximity of Sagui-1, -2, and -3 on the sky, their components are not
artificially merged. Instead, the segmentation responds to differences in morphology, scale, and SNR, effectively disentangling overlapping systems at distinct redshifts. Since our SED fitting procedure assumes a fixed spectroscopic redshift per system, performing it in this case would not be physically meaningful as the three galaxies lie at different redshifts. However, for the other galaxies in our sample, we performed SED fitting on the segmented regions to derive spatially resolved stellar population properties, as shown in Figures~\ref{fig:starlet_sagui4} to \ref{fig:starlet_sagui10}.

\subsection{SED-Fitting}

In this section, we examine the SED-fitting results for the segmented regions of the JADES--Sagui galaxy sample.

\subsubsection{Sagui-4}
For Sagui-4 ($z=0.36$), the algorithm highlights two structures: a bright core, displaced from the galaxy center as a consequence of its disturbed configuration, and two star-forming regions, highlighted by different colours in the segments of the second panel of Figure \ref{fig:mosaic_segmentations}. These star-forming regions may be originally associated with the merged galaxies or they may have formed during the merger, after the fragmentation and compression of the gas. 

\begin{figure}
\centering
\includegraphics[width=\linewidth]{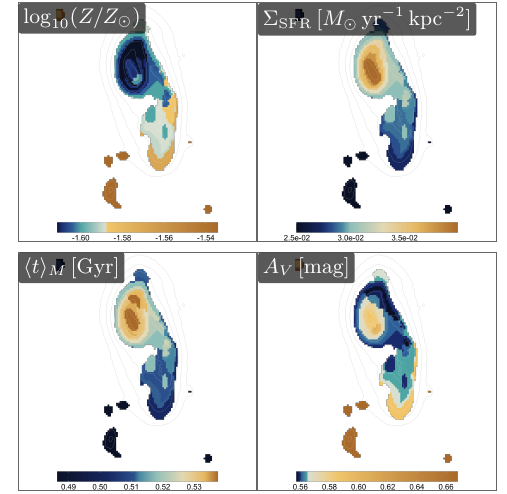}
\caption{Spatially resolved stellar population maps derived from SED fitting for the Sagui-4 system. Each panel displays a two-dimensional map of a physical property inferred from the SED-fitting, including stellar metallicity $\log_{10}(Z/Z_\odot)$, star-formation-rate surface density $\Sigma_{\mathrm{SFR}}\,[M_\odot\,\mathrm{yr}^{-1}\,\mathrm{kpc}^{-2}]$, mass-weighted stellar age $\langle t\rangle_{\mathrm{M}}$, and dust attenuation $A_V$.}
\label{fig:starlet_sagui4}
\end{figure}

Figure~\ref{fig:starlet_sagui4} shows the spatially resolved stellar-population maps derived from the SED fitting for this galaxy. The central region appears to be dominated by an older stellar population, with a relatively small amount of dust, although residual degeneracies cannot be entirely ruled out, also given that we do not include an AGN component in the SED fitting. By contrast, the outer regions show younger stellar populations, with emphasis on some regions that exhibit higher star-formation rates, which coincide with the locations of star-forming clumps visible in the original image of the galaxy (Figure~\ref{fig:jades}). These regions also show modest variations in metallicity and dust attenuation relative to their surroundings. These patterns are consistent with star formation triggered by the merger. While it is beyond the scope of this work, it is important to note here that SED fitting with parametric SFH forms may not accurately capture the complex SFHs of interacting galaxies (e.g. \citealt{2015Smith}). This limitation can, in principle, be mitigated by using SED templates derived from more realistic SFHs obtained from cosmological simulations or semi-analytical models.

\subsubsection{Sagui-5 and -6}

Sagui-5 (face-on) and Sagui-6 (almost edge-on) are a pair of galaxies at the same redshift ($z=0.36$) with asymmetries in their discs probably resulting from tidal perturbations in the early stages of interaction. In Figure \ref{fig:mosaic_segmentations}, the spiral and asymmetric arms of Sagui-5 are prominent, and in Sagui-6, the segmentation clearly shows its elongated disc and larger central component. 

\begin{figure}
\centering
\includegraphics[width=\linewidth]{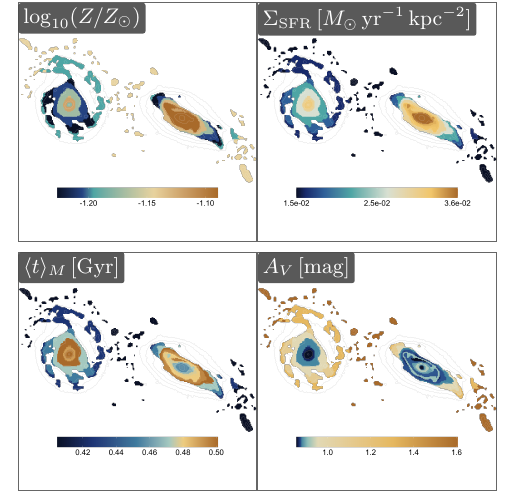}
\caption{Spatially resolved stellar population maps derived from SED fitting for the Sagui-5 and Sagui-6 pair. Each panel displays a two-dimensional map of a physical property inferred from the SED-fitting, including stellar metallicity $\log_{10}(Z/Z_\odot)$, star-formation-rate surface density $\Sigma_{\mathrm{SFR}}\,[M_\odot\,\mathrm{yr}^{-1}\,\mathrm{kpc}^{-2}]$, mass-weighted stellar age $\langle t\rangle_{\mathrm{M}}$, and dust attenuation $A_V$.}
\label{fig:starlet_sagui56}
\end{figure}

In Figure \ref{fig:starlet_sagui56}, the $\Sigma_{\mathrm{SFR}}$ and $A_V$ panels show younger and dustier populations in the spiral arms, particularly in the region disturbed by the ongoing interaction. There is a clear spatial gradient of these properties extending from the inner disc to the outer regions of the galaxies. It is also possible to observe a difference between the two galaxies in terms of metallicity: Sagui-6 has an extended inner region with higher metallicity than Sagui-5.

\subsubsection{Sagui-7}

For Sagui-7 ($z=1.09$), a system of two merging galaxies that still preserve much of their individual structures, the method distinguishes the center of each galaxy and their spiral arms. 

\begin{figure}
\centering
\includegraphics[width=\linewidth]{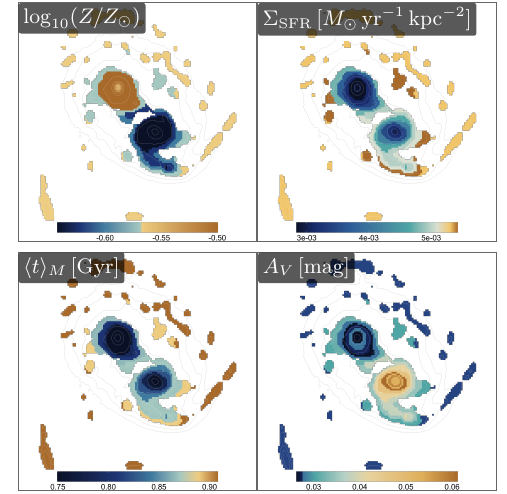}
\caption{Spatially resolved stellar population maps derived from SED fitting for the Sagui-7 system. Each panel displays a two-dimensional map of a physical property inferred from the SED-fitting, including stellar metallicity $\log_{10}(Z/Z_\odot)$, star-formation-rate surface density $\Sigma_{\mathrm{SFR}}\,[M_\odot\,\mathrm{yr}^{-1}\,\mathrm{kpc}^{-2}]$, mass-weighted stellar age $\langle t\rangle_{\mathrm{M}}$, and dust attenuation $A_V$.}
\label{fig:starlet_sagui7}
\end{figure}

Figure \ref{fig:jades} shows that the colliding galaxies have some differences: one of them has more dust (the upper galaxy), while the other is bluer and appears to have more star formation in clumps (the lower galaxy). Differences between the central regions of galaxies also appear in the metallicity and attenuation plots in Figure \ref{fig:starlet_sagui7}. However, since AGN emission was not explicitly modelled, some of the measured attenuation could be affected by residual AGN contribution.

The cores exhibit lower star formation rates than the spiral arms, and it is also interesting to note the increase in star formation rate density near the regions where the collision is occurring (brown regions), suggesting that the \texttt{sagui} segmentation preserves spatially coherent variations in stellar populations.

\subsubsection{Sagui-8}
\begin{figure}
\centering
\includegraphics[width=\linewidth]{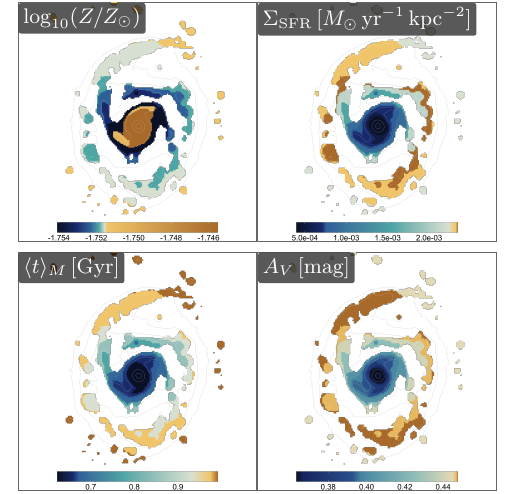}
\caption{Spatially resolved stellar population maps derived from SED fitting for the Sagui-8 system. Each panel displays a two-dimensional map of a physical property inferred from the SED-fitting, including stellar metallicity $\log_{10}(Z/Z_\odot)$, star-formation-rate surface density $\Sigma_{\mathrm{SFR}}\,[M_\odot\,\mathrm{yr}^{-1}\,\mathrm{kpc}^{-2}]$, mass-weighted stellar age $\langle t\rangle_{\mathrm{M}}$, and dust attenuation $A_V$.}
\label{fig:starlet_sagui8}
\end{figure}

Sagui-8 ($z=0.62$) is an undisturbed galaxy with an organized spiral pattern. The \texttt{sagui} segmentation is efficient in separating the well-defined and symmetric arms, highlighting the regions that have different values in their properties compared to their immediate surroundings, as can be seen in the panels of Figure \ref{fig:starlet_sagui8}.

\subsubsection{Sagui-9}

\begin{figure}
\centering
\includegraphics[width=\linewidth]{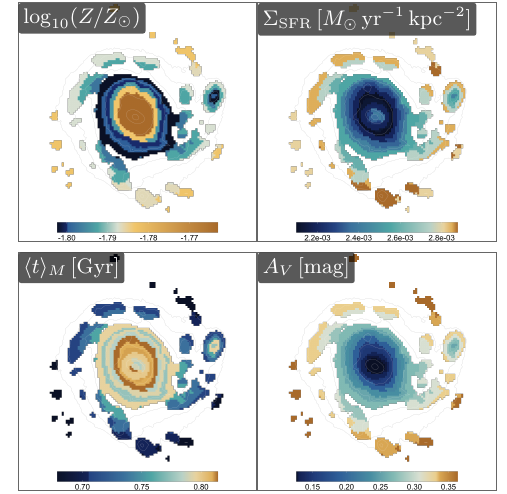}
\caption{Spatially resolved stellar population maps derived from SED fitting for the Sagui-9 system. Each panel displays a two-dimensional map of a physical property inferred from the SED-fitting, including stellar metallicity $\log_{10}(Z/Z_\odot)$, star-formation-rate surface density $\Sigma_{\mathrm{SFR}}\,[M_\odot\,\mathrm{yr}^{-1}\,\mathrm{kpc}^{-2}]$, mass-weighted stellar age $\langle t\rangle_{\mathrm{M}}$, and dust attenuation $A_V$.}
\label{fig:starlet_sagui9}
\end{figure}

The segmentation of Sagui-9 ($z=0.66$) highlights its spiral pattern, as well as a particularly elongated structure in the edge of the disc. This structure is larger than the compact knots that appear in other regions of the disc (see Figure \ref{fig:jades}). This region also stands out in the stellar population maps shown in Figure \ref{fig:starlet_sagui9} because it does not exhibit enhanced star formation compared to the surrounding arm. Previous studies have shown that extended off-centre structures, particularly those larger than typical star-forming clumps, may correspond to accreting satellite galaxies rather than in-situ clumps formed within the disc \citep[e.g.][]{zanella2019}. The location, morphology, and star formation behaviour of this structure may indicate that this is the case here.

\subsubsection{Sagui-10}

Sagui-10 ($z=0.42$) is a strongly barred spiral galaxy. Our segmentation algorithm can separate the arms and the bar, and also distinguish the spherical shape of the central structure. This central component may correspond either to a bulge or to a nuclear disc formed by gas inflows driven by the bar \citep{kormendy2004}, although kinematic information would be required for confirmation. 

\begin{figure}
\centering
\includegraphics[width=\linewidth]{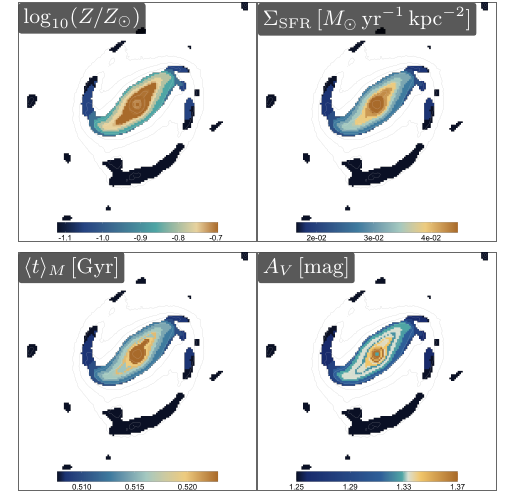}
\caption{Spatially resolved stellar population maps derived from SED fitting for the Sagui-10 system. Each panel displays a two-dimensional map of a physical property inferred from the SED-fitting, including stellar metallicity $\log_{10}(Z/Z_\odot)$, star-formation-rate surface density $\Sigma_{\mathrm{SFR}}\,[M_\odot\,\mathrm{yr}^{-1}\,\mathrm{kpc}^{-2}]$, mass-weighted stellar age $\langle t\rangle_{\mathrm{M}}$, and dust attenuation $A_V$.}
\label{fig:starlet_sagui10}
\end{figure}

Figure \ref{fig:starlet_sagui10} shows that the bar hosts an older stellar population and lower star formation activity compared to the spiral arms. However, the bar is not a uniform structure. The central spheroid and the surrounding bar exhibit spatial variations, mainly in dust attenuation and in the mass-weighted stellar age.

\subsection{The case of low-surface brightness}
\label{sec:lsb}

\subsubsection{Sagui-11}

Sagui-11 is a system of two galaxies at $z = 1.093$ (the spectroscopic redshift of this galaxy was taken from the \citealt{2024Merlin} catalogue). They are connected by a diffuse bridge whose surface brightness lies only marginally above the background level, which provides a challenging and illustrative example for testing the limits of segmentation methods in the presence of LSB emission. 
Such tidal features are common in interacting systems, yet they are notoriously difficult to
recover reliably at intermediate and high redshift, as they are easily suppressed by noise filtering or absorbed
into the background during preprocessing.

In this case, the LSB bridge is morphologically extended but spectrally
coherent, sharing a similar SED shape across multiple \textit{JWST} bands. This
combination makes the system particularly suitable for testing approaches that
go beyond simple surface-brightness thresholding. Methods that rely primarily
on flux contrast or single-band detection tend to fragment or erase such
features, whereas a successful segmentation must simultaneously preserve faint
morphology and inter-band consistency.

\begin{figure}
     \centering
    \includegraphics[width=\linewidth]{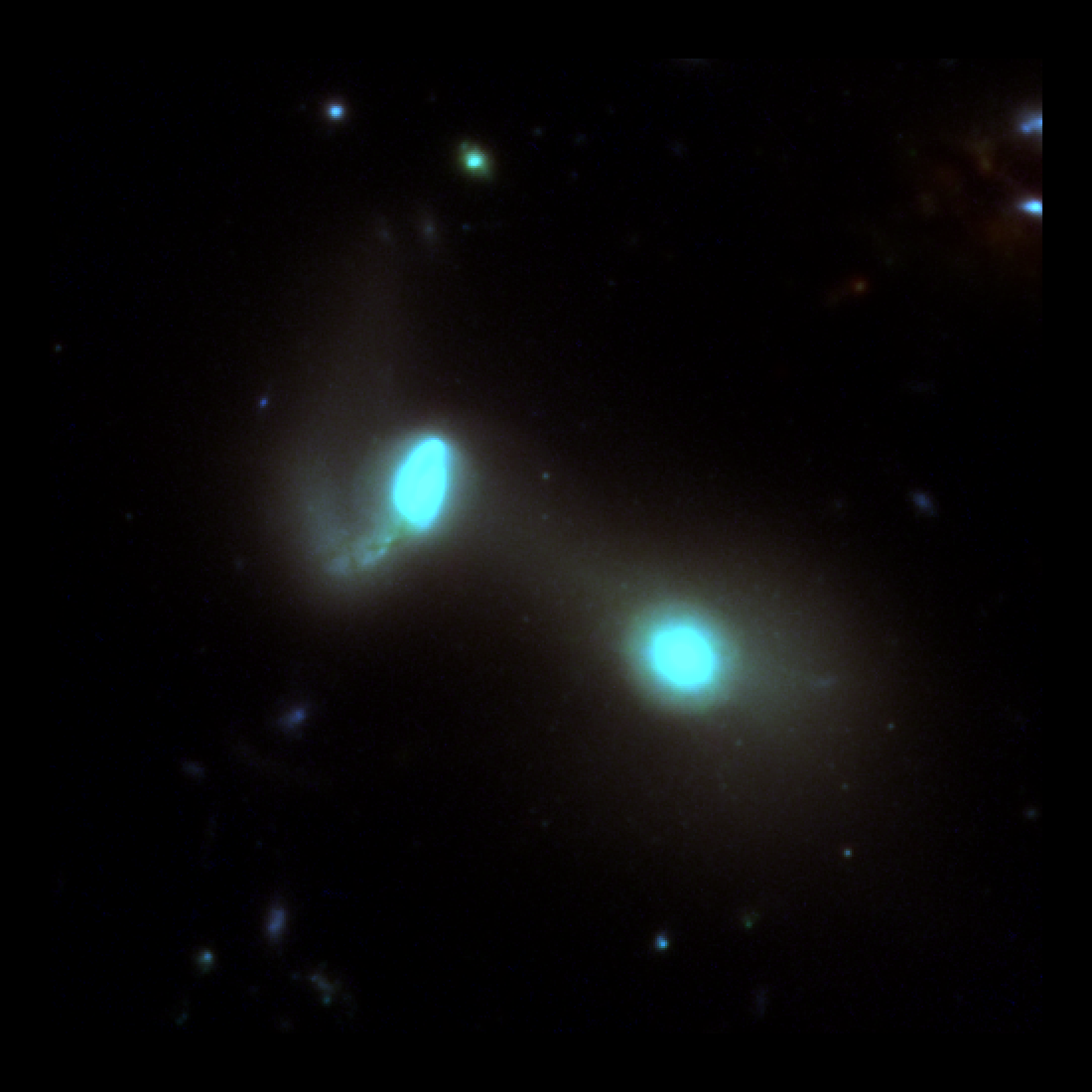}  
    \caption{Interacting galaxies at $z \simeq 1.1$. These galaxies have structures that are examples of low-surface-brightness features: tidal tails and a bridge between them. The image was constructed using the {\it JWST}/NIRCam filters F277W (red), F182M (green) and F115W (blue). It has $\sim$250 pixels per side, corresponding to $\sim$0.167 arcmin.}
    \label{fig:sagui12}
\end{figure}

Figure~\ref{fig:sagui12} shows the composite image of the system, where the
bridge is visually discernible only after aggressive stretching and remains
close to the noise level in individual bands. This regime highlights the need
for a detection strategy that is insensitive to absolute flux scaling, yet
capable of amplifying correlated low-level emission across bands before the
application of the starlet-based mask.

\subsubsection{Copula transform}

To disentangle galaxy light from noisy backgrounds in multi-band imaging, it is essential to understand not only the spatial correlation of flux within each band but also the correlation across different bands. Moreover, this correlation should be characterised in a scale-independent way, since low-surface-brightness galaxies are just as important to segment as their high-surface-brightness counterparts. 

Although galaxies can be segmented using flux density as a metric for spatial correlation, this approach discards valuable information that a copula-based framework can preserve, specifically, scale invariance by removing marginal distributions \citep{nelsen_2006}. By modelling the dependence structure through a copula, we can analyse the spatial and inter-band correlations in a manner that is independent of absolute brightness or size, allowing us to better identify galaxies across a wide dynamic range.

A copula characterises the dependence structure between random variables independently of their marginal distributions. Statistically, it is defined as follows. Let $X_1, \dotsc, X_d$ be random variables with marginal cumulative distribution functions (CDFs) $F_1, \dotsc, F_d$ and joint CDF $F$. Then, by Sklar's theorem, there exists a copula $C$ such that
\begin{equation}\label{eq:sklar}
    F(x_1, \dotsc, x_d) = C(F_1(x_1), \dotsc, F_d(x_d)), \; \; (x_1, \dotsc, x_n) \in \mathcal{R}^d
\end{equation} Here $C$ is a $d$-dimensional CDF with uniform margins on $[0, 1]$ given by $C:[0, 1]^d \to [0, 1]$. This comes from the probability integral transformation $U = F(X)$ and quantile transformation $X = F^{-1}(U)$, where $U$ is uniform on $\mathbf{I}$.

Copulas have found a range of applications in astronomy. They have been used, for example, to construct likelihood functions in large-scale structure \citep{benabed_2009, scherrer_2010} and weak-lensing analyses \citep{sato_2010, Sato2011,Lin2016}, to infer bivariate luminosity and mass functions \citep{Andreani2018}, and to perform multiple imputation of missing data \citep{SPICY2021,Chies2022}. For a more  pedagogical introduction, see also \citet{Feldmann2019,patil_2023}.
For a more comprehensive treatment from a statistical and probabilistic perspective, see \cite{nelsen_2006}. In our context, each image band represents a dimension, with pixel fluxes serving as the random variables whose inter-band dependence we seek to model. To build a single detection image from a multi-band cube that is robust to
band-dependent scalings and nonlinearities, we construct a copula-based
``energy'' map. 

In the present application, the copula is used operationally through the empirical marginal transforms rather than by explicitly fitting a parametric
form for $C$.  Each pixel is mapped from raw multi-band flux space to
empirical copula space by replacing its flux in each band with its marginal
rank, or empirical CDF value. Thus, a pixel \((i,j)\) is represented by the
copula-coordinate vector
\begin{equation}
    \mathbf{U}(i,j)=\bigl(U_1(i,j),\ldots,U_K(i,j)\bigr),
\end{equation}
whose components have uniform margins by construction.
Let $\{I_k\}_{k=1}^K$ denote the $K$ registered bands with
intensities $I_k:\Omega\!\to\!\mathbb{R}$ on the image domain $\Omega$.
For each band, we form an empirical cumulative distribution function
$\hat F_k$ from all pixels in that band, and map intensities to
\emph{normal scores} via the probability integral transform:
\begin{equation}
U_k(i,j) \;=\; \hat F_k\!\big(I_k(i,j)\big), 
\qquad
Z_k(i,j) \;=\; \Phi^{-1}\!\big(U_k(i,j)\big),
\end{equation}
where $\Phi^{-1}$ is the standard normal quantile function. This is equivalent to replacing the uniform margins of the empirical image
copula by standard normal margins, while preserving the rank-based dependence structure across bands. This rank-based step removes per-band monotone transformations, making the construction
insensitive to relative calibration and dynamic-range differences.

For emission-dominated targets (galaxies, spiral arms), positive
departures are most informative. We therefore define a one-sided set of
scores,
\begin{equation}
Z_k^{+}(i,j) \;=\; \max\!\big(Z_k(i,j),\,0\big),
\end{equation}
and aggregate them into a scalar energy via a sum of squares. The resulting
quantity \(P(i,j)\) is therefore not the copula itself, but the positive-tail
squared norm of the pixel in Gaussianized empirical copula space:
\begin{equation}
P(i,j) \;=\; \sum_{k=1}^{K} \big(Z_k^{+}(i,j)\big)^{2}.
\label{eq:copula-energy}
\end{equation}
If the $Z_k$ were independent $\mathcal{N}(0,1)$, then
$\sum_k Z_k^2$ would be $\chi^2_K$; the use of $Z_k^{+}$ concentrates the
statistic on the upper tail, improving sensitivity to faint, positive features while remaining robust to per-band scaling.

\begin{figure*}
\centering
\includegraphics[width=0.95\linewidth]{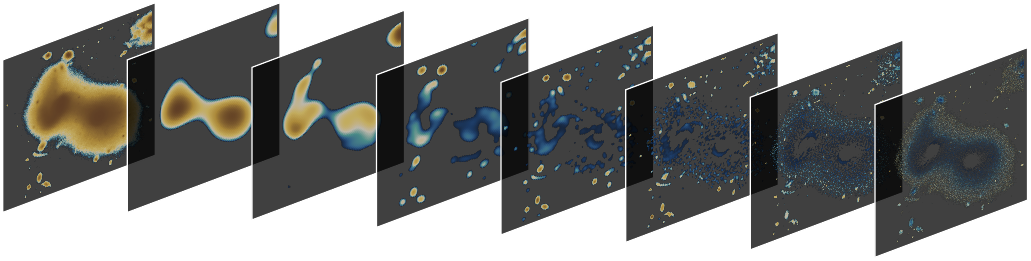}
\caption{Starlet decomposition for representative galaxies (Sagui-11). In each case, the original image is exactly recovered by summing the detail coefficients across all five scales and the coarse component. The finest scale ($j=1$) is dominated by pixel-scale fluctuations, while intermediate scales trace coherent galactic structure. Larger scales isolate progressively smoother and more extended components.}
\label{fig:starlet_sagui11}
\end{figure*}

\begin{figure}
     \centering
    \includegraphics[width=0.45\linewidth]{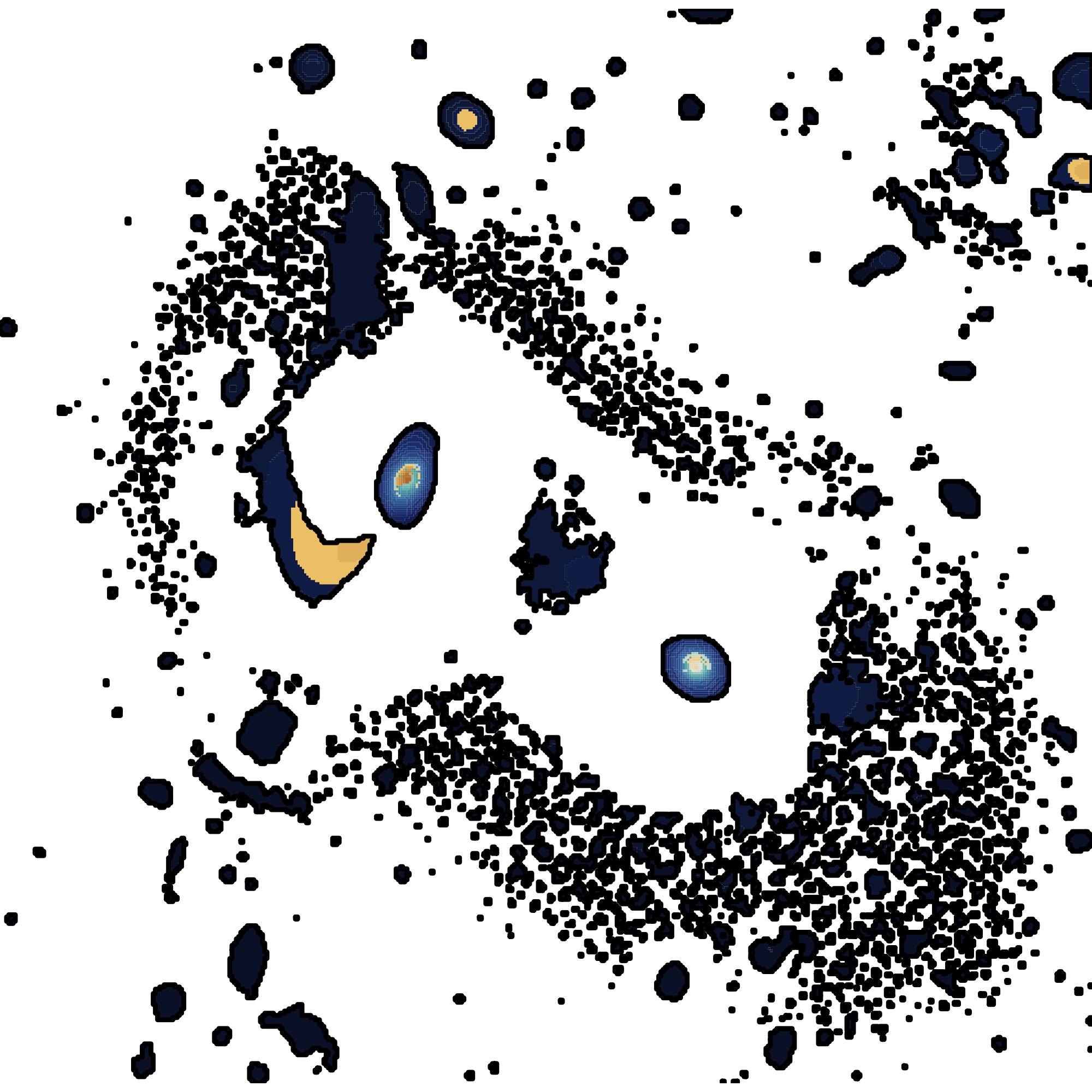} 
       \includegraphics[width=0.45\linewidth]{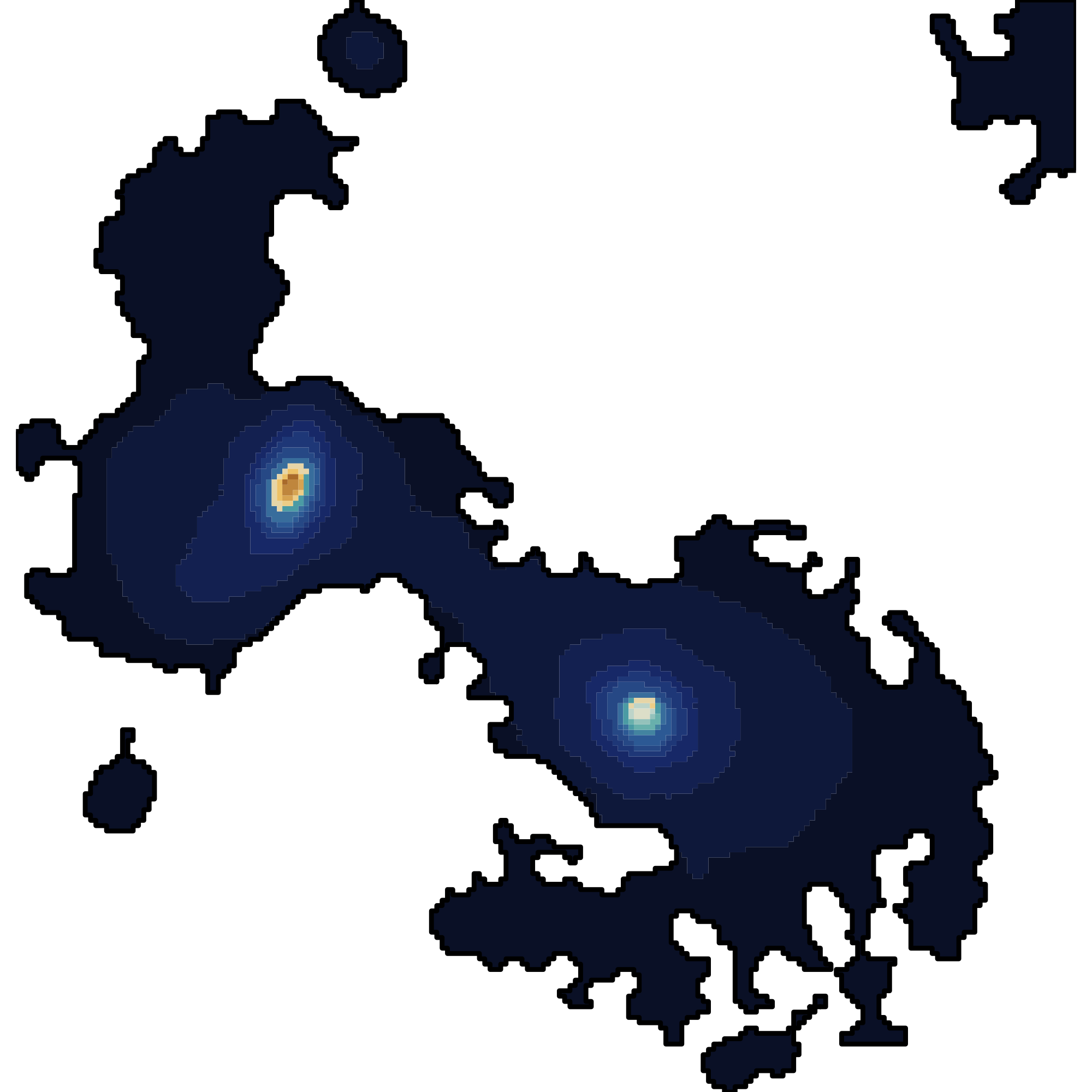}  
    \caption{Spatial distribution of the 20 segments detected by \texttt{sagui}. Different colours denote distinct segments. 
    Left panel: segmentation performed directly on the data, as in Fig.~\ref{fig:mosaic_segmentations}. 
    Right panel: segmentation performed on the copula-transformed data.}
    \label{fig:segments_sagui12}
\end{figure}

\begin{figure}
\centering
\includegraphics[width=\linewidth]{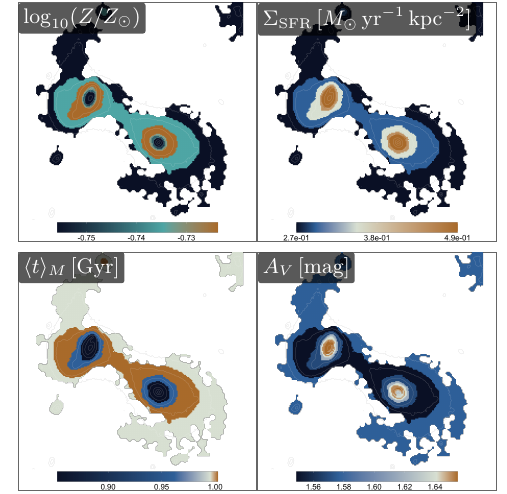}
\caption{Spatially resolved stellar population maps derived from SED fitting for the Sagui-11 system. Each panel displays a two-dimensional map of a physical property inferred from the SED fitting, including stellar metallicity $\log_{10}(Z/Z_\odot)$, star-formation-rate surface density $\Sigma_{\mathrm{SFR}}\,[M_\odot\,\mathrm{yr}^{-1}\,\mathrm{kpc}^{-2}]$, mass-weighted stellar age $\langle t\rangle_{\mathrm{M}}$, and dust attenuation $A_V$.}
\label{fig:map_sagui11}
\end{figure}

Figure~\ref{fig:starlet_sagui11} shows the starlet decomposition for this pair of galaxies after applying copula, while Figure~\ref{fig:segments_sagui12} shows the segmentation using \texttt{sagui}. As shown by the two panels, the subtle LSB features would be missed if the copula-based analysis were not applied. The two galaxies are clearly visible; unlike the previous examples, they do not have many internal structures, as they are both red galaxies, but they do have an external bridge connecting them due to their interaction. In addition, the northern galaxy has a very prominent tidal tail, and the southern galaxy appears more hexagonal in shape due to the PSF shape at wider filters. 
In Figure~\ref{fig:map_sagui11}, the bridge also stands out in terms of the properties shown: on the $\Sigma_{\mathrm{SFR}}$ map, naturally, the star formation rate is low in this structure compared to galaxies. This is also true for dust attenuation and metallicity. These results demonstrate that combining \texttt{sagui} with copula-based analysis enables a coherent mapping of both spatial and physical properties, providing insights into galaxy morphology, and LSB features.

\section{Conclusions}
\label{sec:conclusions}

We present a framework for the analysis of multi-band imaging data, implemented in our package \texttt{sagui}: a modular approach for spatially resolved galaxy analysis. Building on the spectro--spatial paradigm introduced by \textsc{capivara} for IFS data, \textsc{sagui} extends this philosophy to multi-band imaging, enabling a coherent pixel-level treatment of spatial and spectral information across multiple filters. The method follows a two-stage strategy. In the first stage, a starlet-based decomposition is used to identify and mask spatial structures across multiple scales while suppressing background noise. In the second stage, a spectral similarity analysis partitions the image into coherent pixel groups that preserve spectral consistency across bands.

We demonstrate the performance of the method across a diverse range of galaxy morphologies, illustrating its ability to recover complex spatial structures such as clumps, bars, and interacting systems. As a case study, we apply the framework to eleven morphologically diverse galaxies from  JADES in the GOODS--South field. 

By construction, the method is not intended as a replacement for source-detection or deblending pipelines, but rather as a complementary segmentation layer designed for resolved photometric analysis. Its main strength lies in combining a morphology-aware spatial support with a segmentation performed in SED space, allowing the identification of regions that are photometrically homogeneous. This is particularly relevant in situations where conventional geometric binning may improve S/N ratio at the cost of mixing distinct stellar populations or blurring substructure.

The application to the JADES sample shows that this strategy can recover meaningful internal organization across a broad range of systems, from relatively undisturbed spirals to strongly asymmetric and interacting galaxies. When combined with region-based SED fitting, the segmentation provides an effective route toward spatially resolved maps of stellar population properties, while retaining a close connection to the observed morphology. In this sense, \textsc{sagui} provides a practical interface between modern multi-band imaging surveys and downstream inference tools aimed at characterizing the internal physics of galaxies.

A current limitation of the method is that spatial contiguity is not explicitly enforced during clustering. Since the segmentation is driven primarily by similarity in SED space, spatially disconnected regions may be assigned to the same \textsc{sagui} group when their multi-band photometric properties are sufficiently similar. For SED-based spatial maps this is not necessarily problematic, as physically distinct regions may share similar colours or stellar-population properties. However, such segments should not always be interpreted as spatially contiguous morphological units. When needed, contiguity can be introduced through a post-processing layer, such as connected-component or spatial-instance labels within each \textsc{sagui} group, or more directly through spatially regularized clustering.

Looking ahead, the framework is naturally suited to extensions in both methodology and application. On the methodological side, future developments will explore alternative similarity measures tailored to specific scientific goals, including studies of emission lines, kinematics, and related observables, as well as more flexible clustering strategies, spatial regularization, and improved treatments of diffuse low-surface-brightness components. These extensions may be particularly relevant where kinematics and local continuity often carry direct physical meaning. On the observational side, the approach is readily applicable to forthcoming large imaging datasets, for which physically informed and scalable segmentation will become increasingly important. We therefore view \textsc{sagui} as a step toward a more coherent analysis of resolved galaxy structure in the era of deep, multi-band surveys.

\section*{DATA AVAILABILITY}

The \textit{JWST}/JADES imaging data used in this analysis are publicly
available from the \href{https://archive.stsci.edu/}{Mikulski Archive for Space Telescopes (MAST)} as High-Level
Science Products under DOI: \href{https://doi.org/10.17909/8tdj-8n28}{10.17909/8tdj-8n28}.
Part of the data products presented here were retrieved from the Dawn
\textit{JWST} Archive (DJA), an initiative of the Cosmic Dawn Center (DAWN),
which is funded by the Danish National Research Foundation under grant DNRF140.

\section*{Acknowledgements}
This work was developed as part of CRP8, which was partially supported by the CNPq grant 445231/2024-6. SS acknowledges support from the UK Science and Technology Facilities Council (STFC) via the grant ST/X508408/1.
ACS acknowledges support from FAPERGS (grants 23/2551-0001832-2 and 24/2551-0001548-5), CNPq (grants 314301/2021-6, 312940/2025-4, 445231/2024-6, and 404233/2024-4), and CAPES (grant 88887.004427/2024-00).  This work is based in part on observations made with the NASA/ESA/CSA \textit{James Webb Space Telescope}. The data were obtained from the Mikulski
Archive for Space Telescopes at the Space Telescope Science Institute, which is
operated by the Association of Universities for Research in Astronomy, Inc.,
under NASA contract NAS 5-03127 for JWST. These observations are associated
with the JWST Advanced Deep Extragalactic Survey (JADES).




\bibliographystyle{mnras}
\bibliography{references} 




\appendix

\section{1D Toy Example of the Starlet Reconstruction}

To make the multiscale structure of the starlet transform more explicit, this appendix presents a simple one-dimensional toy example. Our goal here is purely pedagogical: to illustrate how the \`a trous construction generates a sequence of progressively dilated B-spline kernels, how these kernels produce smoother approximations of the input signal at increasing scales, and how the associated detail coefficients encode the structures resolved at each level. We first show the filter kernels themselves and then the resulting decomposition of the toy signal into smooth and detail components.
 
\begin{figure}
     \centering
     \includegraphics[width=\linewidth]{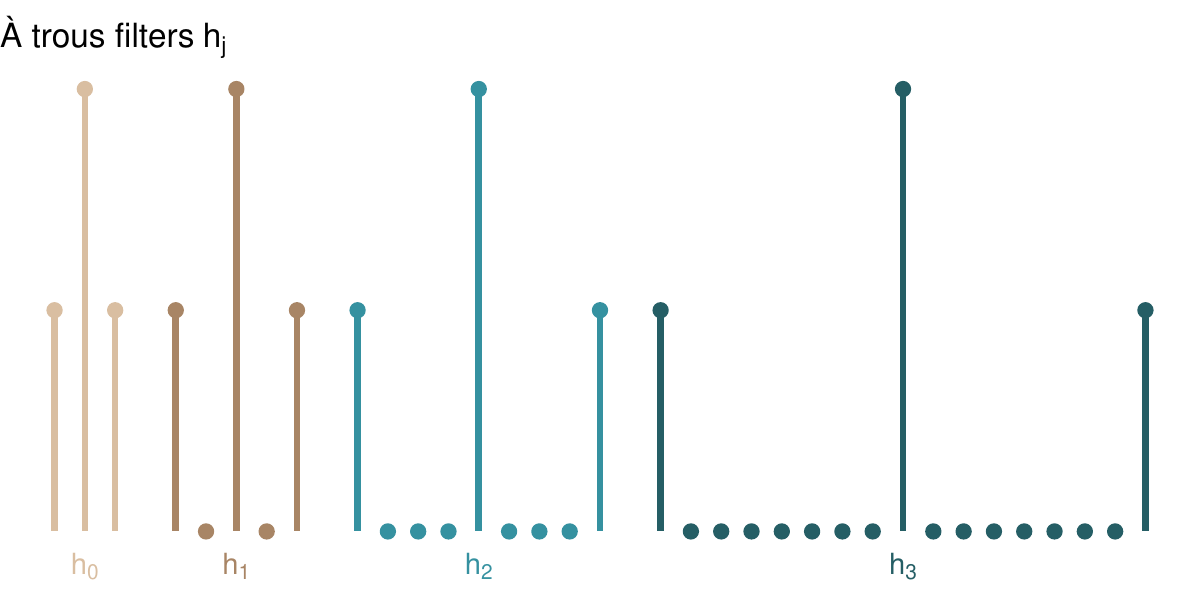}
    \caption{
\`A trous filter kernels \(h_j\) for a toy model, shown for successive scales
\(j=0,\ldots,J-1\). The widening support with increasing \(j\) illustrates the
standard dilation-by-holes construction applied to the base filter \(h_0\).
The horizontal index is arbitrary and is used only to visually separate the
kernels.
}
 \label{fig:starlet_kernel}
 \end{figure}

\begin{figure}
     \centering
     \includegraphics[width=\linewidth]{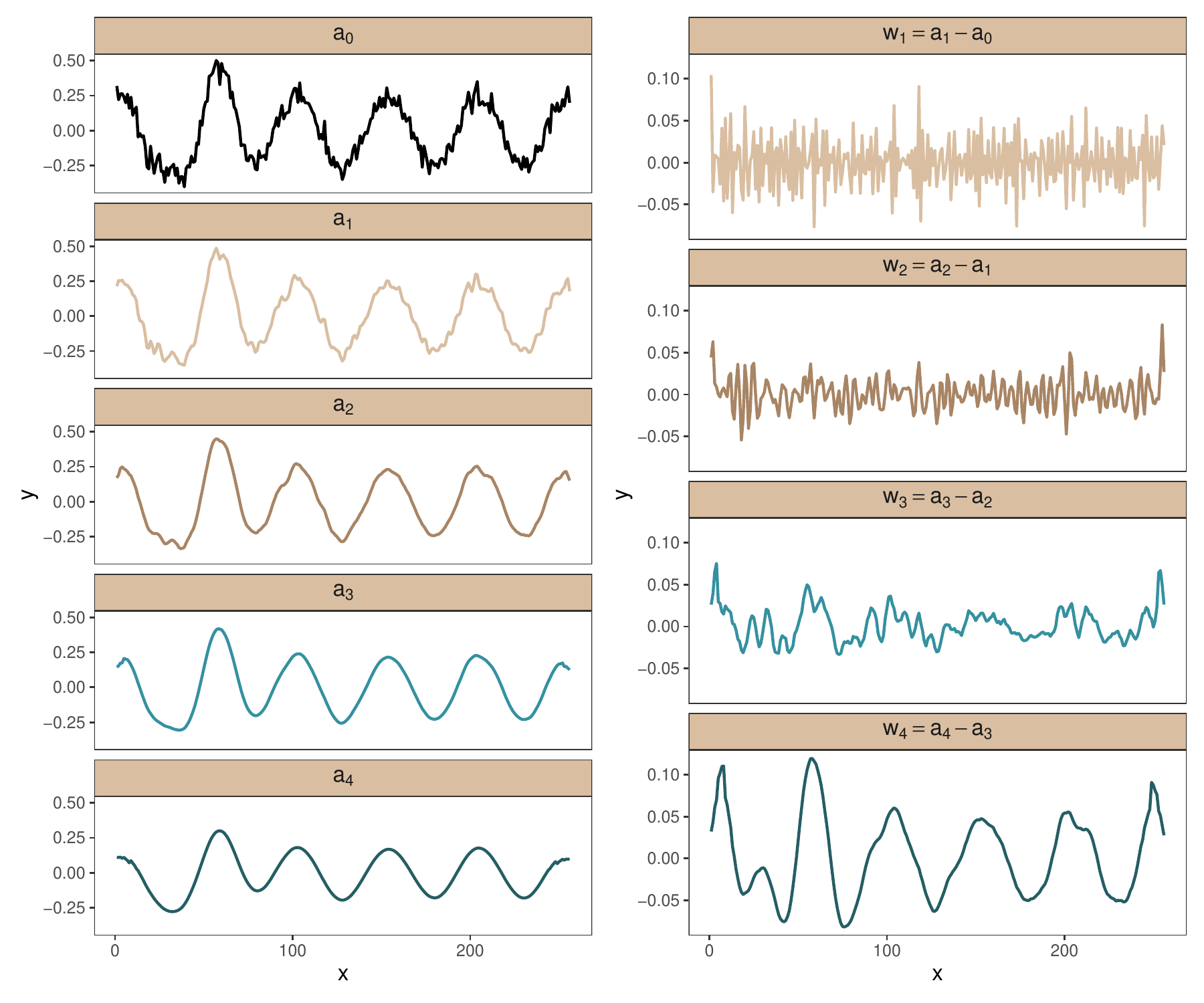}
 \caption{
Starlet (\`a trous) decomposition of a one-dimensional signal into its smooth components \(a_j\) (left column) and detail coefficients
\(w_j = a_{j-1} - a_j\) (right column). The sequence \(\{a_j\}\) provides
progressively smoother approximations to the original function (\(a_0\)),
while \(\{w_j\}\) captures the fluctuations resolved at each spatial scale.
The sum of all detail components together with the coarsest smooth component
reconstructs the original function to numerical precision.}
 \label{fig:starlet_components}
 \end{figure}

We decompose the input function using the isotropic starlet transform, which
iteratively convolves the signal with the dilated B--spline scaling kernels
\(h_j\) shown in Figure~\ref{fig:starlet_kernel}.  This procedure yields a sequence of smooth components \(a_j\) and
corresponding detail coefficients \(w_j = a_j - a_{j-1}\). As the effective
support of \(h_j\) increases with \(j\), the smooth terms \(a_j\) provide
progressively coarser approximations of the underlying function, while the
detail components \(w_j\) isolate the structures introduced at each scale.  
Because the starlet transform is redundant and exactly invertible, the original
function may be reconstructed---up to numerical precision---by summing the
coarsest smooth component and all detail coefficients,
\[
f(x) = a_J(x) + \sum_{j=1}^{J} w_j(x).
\]
Figure~\ref{fig:starlet_components} illustrates this multiscale decomposition
for our toy model: large-scale behaviour accumulates in the coarse
approximation \(a_J\), whereas fine-scale oscillatory variations are confined
to the wavelet coefficients \(w_j\).

\section{Starlet Decomposition}
\begin{figure}
\centering
\includegraphics[width=0.9\linewidth]{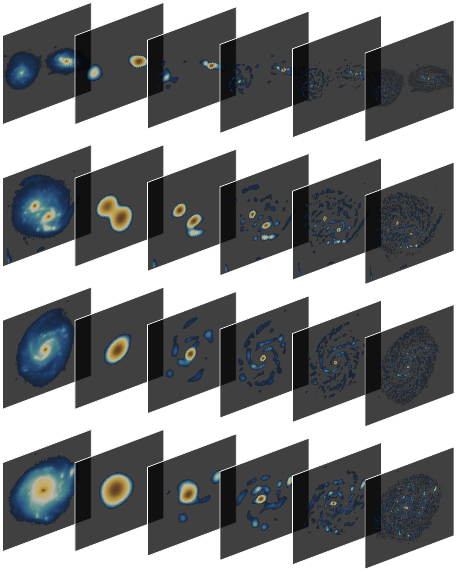}
\caption{Starlet decomposition for representative galaxies (Sagui-5–9). In each case, the original image is exactly recovered by summing the detail coefficients across all five scales and the coarse component. The finest scale ($j=1$) is dominated by pixel-scale fluctuations, while intermediate scales trace coherent galactic structure. Larger scales isolate progressively smoother and more extended components.}
\label{fig:starlet_sagui59}
\end{figure}

Here in Figure~\ref{fig:starlet_sagui59} we show the 2D maps of the starlet components for the other galaxies in our sample. 

\bsp	
\label{lastpage}
\end{document}